\documentclass[twocolumn,aps,pre,reprint,url = false, showkeys, longbibliography,
groupaddress,
,en,footinbib]{revtex4-2}
\usepackage[utf8]{inputenc}
\usepackage[english]{babel}
\usepackage{amssymb}
\usepackage{amsmath}
\usepackage{comment}
\usepackage{lipsum}
\usepackage{bm}
\usepackage{booktabs}
\usepackage{hyperref}

\usepackage{mhchem}

\usepackage{mathtools}
\usepackage[final]{graphicx}

\usepackage[labelformat=simple,subrefformat=parens,labelformat=parens]{subfig}
\usepackage{floatrow}
\usepackage{caption}
\newcommand{\RN}[1]{%
  \textup{\uppercase\expandafter{\romannumeral#1}}%
}

\usepackage[usenames, dvipsnames]{color}

\definecolor{myred1}{RGB}{255, 0, 0}
\definecolor{myyellow1}{RGB}{255, 255, 219}
\definecolor{mygreen1}{RGB}{0, 255, 0}
\definecolor{mygreen2}{RGB}{0, 126, 0}
\definecolor{myblue1}{RGB}{0, 0, 255}
\usepackage{physics}

\begin{document}
\title{
    %
    How Fluctuations Control Chiral Symmetry Breaking in \\
    Reacting Mixtures of Chiral Molecules
}

\author{Leonardo Silva-Dias}
\affiliation{
 Faculty of Mathematics, Natural Sciences, and Engineering: Institute of Physics, University of Augsburg, Universit\"atsstra\ss e~1, 86159 Augsburg, Germany
}



\author{Christoph A.\,Weber}

\affiliation{
 Faculty of Mathematics, Natural Sciences, and Engineering: Institute of Physics, University of Augsburg, Universit\"atsstra\ss e~1, 86159 Augsburg, Germany
}


\begin{abstract}
The pronounced chiral asymmetry of biomolecules is a defining feature of living systems, yet the mechanisms by which such asymmetry emerged remain unresolved. Chiral symmetry breaking may have occurred under prebiotic conditions in systems with sufficiently small molecular populations for intrinsic fluctuations to play an important role. Here, we investigate how finite-population fluctuations affect chiral-state selection, the characteristic time for symmetry breaking, and stochastic switching between chiral states in a system driven by autocatalysis and mutual antagonism. We determine how these dynamical processes scale with particle number and derive an analytical expression for the probability of selecting either chiral state as a function of the initial enantiomeric bias. We further calculate the mean selection time for different initial biases and show that, for weak mutual antagonism, fluctuations can induce transitions between opposite chiral states. These results characterize how finite-population fluctuations control both the emergence and stability of molecular chirality in autocatalytic chemical systems.

\end{abstract}
\maketitle

\section{Introduction}

A striking feature of biological systems is the pronounced chiral asymmetry of their molecular building blocks, with amino acids and carbohydrates occurring predominantly in specific enantiomeric forms \cite{hegstrom1990handedness,saito2013colloquium}. The physicochemical mechanisms by which this asymmetry emerged and a particular handedness was selected remain incompletely understood. It has been proposed that chiral symmetry breaking occurred spontaneously under nonequilibrium conditions \cite{nicolis1977self,kondepudi1983chiral}, leading to the emergence of an enantiomeric excess from an initially racemic mixture, i.e., a mixture containing equal amounts of the two enantiomers \cite{atkins1999chemical}. This longstanding problem has motivated extensive experimental and theoretical work aimed at identifying mechanisms capable of generating and amplifying chiral asymmetry.

The emergence and amplification of chiral asymmetry have been experimentally demonstrated in a variety of chemical systems, ranging from crystal growth to the synthesis of organic molecules \cite{kondepudi2001chiral,saito2013colloquium}. In particular, chiral symmetry breaking during the crystallization of sodium chlorate under stirring, and subsequently under grinding with beads, provided the basis for methods used to obtain other chiral crystals of inorganic and organic compounds \cite{kondepudi1990chiral,viedma2005chiral,kondepudi1993kinetics,buhse2000chiral,cartwright2004chiral,noorduin2008emergence,viedma2008evolution,tsogoeva2009generation}. In organic chemistry, asymmetric synthesis has been demonstrated in the catalyzed alkylation of aldehydes through the Soai reaction \cite{soai1995asymmetric,sato2001evidence,soai2019asymmetric}. Chiral amplification has also been observed in homochiral dipeptide products formed by a prebiotic ligation reaction \cite{deng2024symmetry}. These experiments have provided important insight into the mechanistic requirements for the emergence, amplification, and selection of chiral asymmetry and have motivated the development of theoretical models.

A famous theoretical model of spontaneous chiral symmetry breaking was proposed by F.~C.~Frank~\cite{frank1953spontaneous}. 
In Frank’s mechanism, a system initially containing equal amounts of the two enantiomers can undergo a supercritical pitchfork bifurcation through the combined effects of autocatalysis and competition. At the bifurcation, the stable racemic state becomes unstable and two symmetry-related stable chiral states emerge \cite{kondepudi1983chiral,gillet2024differences}. In deterministic dynamics, even a small initial bias can then determine which of these states is selected. Frank’s work provided the basis for numerous models incorporating additional features of specific chemical systems \cite{kondepudi1983chiral,kondepudi1984chiral,blackmond2008re,saito2004complete,plasson2004recycling,boscheto2015spontaneous}. Most of these models are deterministic and show how an initial bias, physical perturbations, or spatial effects can influence the selection of a particular handedness. Although deterministic models provide important insight into the mechanisms of chiral symmetry breaking, they neglect fluctuations arising from finite molecular populations. Describing such systems therefore requires stochastic models that explicitly account for intrinsic fluctuations.

Intrinsic fluctuations are unavoidable in chemical systems containing a finite number of molecules. For chiral symmetry breaking, their effects have been investigated mainly in two distinct scenarios: (i) systems in which intrinsic fluctuations themselves drive the transition from a racemic to a chiral state, and (ii) systems in which a deterministic mechanism, such as Frank’s mechanism of autocatalysis and competition, drives chiral symmetry breaking while intrinsic fluctuations influence the selection and subsequent dynamics of the resulting chiral states.

In the first scenario (i), recent studies have shown that complete homochirality can emerge in stochastic models of chiral self-replicators without requiring chiral inhibition~\cite{jafarpour2015noise,jafarpour2017noise}. In these models, autocatalytic reactions produce molecules of the same chirality, while decay reactions recycle them into achiral precursors. The stochasticity of the autocatalytic reactions generates multiplicative noise whose amplitude depends on the chiral composition of the system: fluctuations are strongest near the racemic state and vanish at the homochiral states, thereby driving the system toward and stabilizing it in one of the two homochiral states \cite{jafarpour2015noise,jafarpour2017noise}. Chiral asymmetry can also emerge spontaneously in finite stochastic chemical reaction networks coupled to external energy sources as a consequence of intrinsic fluctuations, even in the absence of an imposed chiral bias or an explicit mechanism involving autocatalysis and chiral inhibition \cite{pineros2022spontaneous}.

In the second scenario (ii), stochastic versions of Frank’s model have numerically demonstrated that, in the absence of an initial bias, intrinsic fluctuations can cause the system to spontaneously select either of the two chiral states, with the selected handedness determined randomly \cite{silva2017stochastic,lente2010role,silva2018spontaneous,hochberg2023chiral}. In the presence of an initial enantiomeric excess, however, intrinsic fluctuations can cause the system to select a chirality different from that favored by the corresponding deterministic dynamics. The initial bias therefore competes with finite-population fluctuations, and the enantiomer initially in excess is not necessarily the one ultimately selected \cite{kondepudi1985weak,silva2017stochastic,lente2010role,sugimori2008fluctuation,de2025stochastic,lente2005stochastic,silva2018spontaneous,lente2009stochastic}. This competition raises several questions that remain poorly understood: How does the probability of selecting either chiral state depend on the initial enantiomeric bias and the number of molecules? How do these quantities determine the characteristic time required for chiral symmetry breaking? Under what conditions can intrinsic fluctuations induce switching between opposite chiral states?

In this work, we investigate how intrinsic fluctuations affect chiral symmetry breaking in finite chemical systems governed by Frank’s mechanism of autocatalysis and competition. Starting from the chemical master equation, we derive a Fokker–Planck equation through a Kramers–Moyal expansion and use it to obtain analytical expressions for the stationary distribution, the splitting probability between the two chiral states, and the mean first-passage times associated with chiral selection and switching. These results allow us to determine how the probability and characteristic time of chiral selection depend on system size and initial enantiomeric excess. In particular, we characterize the competition between an initial chiral bias and finite-population fluctuations and determine how this competition scales with the number of molecules. Finally, we identify the conditions under which intrinsic fluctuations can induce switching between opposite chiral states. Together, these results provide an analytical characterization of how finite-population fluctuations govern both the selection and stability of chiral states in autocatalytic chemical systems.

\section{Model}

\subsection{Kondepudi-Nelson model}

As discussed in the introduction, there exist different types of theories proposed to explain the emergence homochirality. However, the most influential one is based on the hypothesis proposed by F.~C.~Frank in 1953~\cite{frank1953spontaneous}. 
In his seminal paper, Frank suggested and numerically demonstrated that chiral symmetry breaking may emerge in chemical systems in which the enantiomers undergo autocatalysis and mutual competition. The first process amplifies the concentration of molecules with specific chirality, whereas the second corresponds to a mutually antagonistic interaction between the two enantiomers that leads to their destruction. Frank's model served as the foundation for the development of several variants, such as the Kondepudi--Nelson model \cite{kondepudi1983chiral,kondepudi1984chiral}:

\begin{center}

\ce{A <=>[k_1][k_2] X} \, ,

\ce{A <=>[k_1][k_2] Y}\, ,

\ce{B + X <=>[k_{3}][k_{4}] 2X}\, ,

\ce{B + Y <=>[k_{3}][k_{4}] 2Y}\, ,

\ce{X + Y ->[k_5] P}\, .

\end{center}

This model is a simple extension of Frank's original model. It consists of two reactions in which chiral molecules (X and Y) are produced from an achiral substrate (A and B), two autocatalytic reactions, and one competition reaction, which irreversibly forms a product (P). It can be shown that, by controlling specific parameters, the system may undergo a bifurcation, transitioning from a stable racemic state to an unstable one, in which small concentration imbalances between the chemical species can lead to complete symmetry breaking. Throughout this work, we will consider this model and focus particularly on its unstable regime that leads to chiral selection.

\subsection{The Master equation}

In the following, we begin the derivation of the model through the formulation of the Master equation for the Kondepudi--Nelson model. 
We consider a well-mixed and dilute system in which spatial correlations can be neglected. 
The system is in contact with a chemostat that maintains the concentrations of A, B, and P approximately constant throughout the dynamics. 
This feature is essential for
keeping the system out of equilibrium. Under such conditions, the state of the system is defined by the vector
$\mathbf q = (X(t),Y(t))$,
where $X(t)$ and $Y(t)$ denote the number of particles of the enantiomers X and Y, respectively. For a finite system, these quantities evolve discretely through the chemical reaction pathways introduced in the previous section.

To proceed, we define the stoichiometric matrix

\begin{equation} \label{eq:1}
\mathbf S =
\left(
\boldsymbol{\nu}_1,
\boldsymbol{\nu}_2,
\ldots,
\boldsymbol{\nu}_R
\right),
\end{equation}

where $\boldsymbol{\nu}_i$ is the stoichiometric vector associated with reaction $i$. The matrix $\mathbf S$ is therefore an $M\times R$ matrix, where $M$ is the number of chemical species and $R$ is the number of reactions. Each column $\boldsymbol{\nu}_i \in \mathbb Z^M$ specifies the change in the state vector produced by the occurrence of reaction $i$. For the present model, the stoichiometric matrix is given by

\begin{equation}\label{eq:2}
\mathbf S =
\begin{pmatrix}
1 & -1 & 0 & 0 & 1 & -1 & 0 & 0 & -1 \\
0 & 0 & 1 & -1 & 0 & 0 & 1 & -1 & -1
\end{pmatrix}.
\end{equation}

We further define the transition rate $W_i(\mathbf q,t)$ as the probability per unit time that reaction $i$ occurs when the system is in state $\mathbf q$. The transition rates associated with the reaction channels are given by
\begin{equation}\label{eq:3}
\begin{aligned}
W_1(\mathbf q,t) &= k_1 A, \\
W_2(\mathbf q,t) &= k_2 X, \\
W_3(\mathbf q,t) &= k_1 A, \\
W_4(\mathbf q,t) &= k_2 Y, \\
W_5(\mathbf q,t) &= k_3 BX N^{-1}, \\
W_6(\mathbf q,t) &= 2k_4 X(X-1) N^{-1}, \\
W_7(\mathbf q,t) &= k_3 BY N^{-1}, \\
W_8(\mathbf q,t) &= 2k_4 Y(Y-1) N^{-1}, \\
W_9(\mathbf q,t) &= k_5 XY N^{-1}.
\end{aligned}
\end{equation}

With these definitions, the Master equation governing the time evolution of the probability $P(\mathbf q,t)$ of finding the system in state $\mathbf q$ at time $t$ can be written as

\begin{equation}\label{eq:4} 
\begin{split} 
\partial_t P(\mathbf q,t) = \sum_{i=1}^{R} & \left[ W_i(\mathbf q -\boldsymbol{\nu}_i,t) P(\mathbf q-\boldsymbol{\nu}_i,t) \right. \\ 
& \left. - W_i(\mathbf q,t) P(\mathbf q,t) \right]. 
\end{split} 
\end{equation}

 From the Master equation, we derive the dynamical equations governing the evolution of the first moments, which, in the thermodynamic limit, correspond to the deterministic evolution of the system. Moreover, the Master equation allows the derivation of the Fokker--Planck equation, which provides a continuous approximation to the Master equation and constitutes a more tractable mathematical framework for obtaining the probability distribution of the system \cite{frey2010evolutionary,reichenbach2006coexistence}. These derivations are presented in the following sections.

\subsection{Determinsitic rate equations}\label{subsec:IIC}

The deterministic rate equations describe the evolution of the densities
$x(t)= {\langle X(t)\rangle}/{N}$ and $y(t)= {\langle Y(t)\rangle}/{N}$, corresponding to the concentrations of the chemical
species $X$ and $Y$, respectively. These equations can be obtained from
the Master equation (Eq.~(\ref{eq:4})) by computing the first moments,
\begin{subequations}
\begin{align}
x(t) &
= \frac{1}{N}\sum_{\mathbf q} X P(\mathbf q,t), \\
y(t) &=  \frac{1}{N}\sum_{\mathbf q} Y P(\mathbf q,t),
\end{align}
\end{subequations}
and taking the thermodynamic limit of a large number of particles
(or equivalently large volume), which leads to the following deterministic
dynamical equations:

\begin{subequations}\label{eq:5}
\begin{align}
\dot{x} &= k_1 a + x(k_3 b - k_2) - 2k_4 x^2 - k_5 xy, \\
\dot{y} &= k_1 a + y(k_3 b - k_2) - 2k_4 y^2 - k_5 xy.
\end{align}
\end{subequations}
These equations can also be obtained directly from the law of mass action, and they describe the kinetics of the system in the regime where fluctuations are effectively negligible. 

For the purposes of this work, it is convenient to introduce new variables
based on the symmetry of the system:
\begin{equation}
\phi(t) = x(t) + y(t),
\qquad
\psi(t) = x(t) - y(t).
\end{equation}
Here, $\phi(t)$ represents the total concentration of the chiral species,
while $\psi(t)$ measures the chiral imbalance of the system. The dynamical
equations for these new variables, obtained from Eqs.~(\ref{eq:5}), are
\begin{subequations}\label{eq:6}
\begin{align}
\dot{\phi}(\tau) &=
2\delta + (\sigma - \gamma)\phi
- \frac{\omega+\epsilon}{2}\phi^2
+ \frac{\epsilon-\omega}{2}\psi^2,\\
\dot{\psi}(\tau)
&=
\psi\left[(\sigma-\gamma)-\omega\phi\right].
\end{align}
\end{subequations}
Above, the time variable in Eqs.~(\ref{eq:6}) has been rescaled according to the dimensionless time $\tau = k_1 t$. We also introduced the redefined parameters $\delta = a$, $\gamma = {k_2}/{k_1}$, $\sigma = {k_3 b}/{k_1}$, $\omega = {2k_4}/{k_1}$, and $\epsilon = {k_5}/{k_1}$.

\subsubsection{Fixed points and linear regime}

The nonlinear nature of Eqs.~(\ref{eq:6}) generally requires numerical methods for a complete characterization of the dynamics. Nevertheless, important analytical insights can be obtained by studying the linearized equations in the vicinity of the fixed points \cite{silva2021basic}. 

Assuming all parameters, i.e., $\delta,~\gamma,~\sigma,~\omega,~ \text{and} ~\epsilon$, are positive, the following conditions guarantee the existence of one symmetric and two asymmetric real and positive fixed points ($(\dot{\psi}, \dot{\phi})=(0,0)$):

\begin{equation*}
    \omega \neq 0,
    \qquad
    \epsilon > \omega,
    \qquad
    \sigma > \gamma,
    \qquad
    \frac{(\sigma-\gamma)^{2}}{\omega^{2}}
    +
    \frac{4\delta}{\omega-\epsilon}
    > 0.
\end{equation*}

Since our main interest is the spontaneous breaking of chiral symmetry, we focus throughout this work on the parameter regime defined by the conditions above. Under these assumptions, the behavior of the trajectories near the fixed points can be analyzed through linear stability analysis, the details of which are presented in Appendix \ref{subsec:AppA}. This analysis shows that the system possesses two stable attractive asymmetric fixed points and one unstable symmetric saddle point. In physical terms, the symmetric (racemic) state is unstable, whereas the asymmetric (chiral) states are locally stable, configuring a bistability regime.

\begin{figure}[t]
    \centering
    \includegraphics[width=1.0\linewidth]{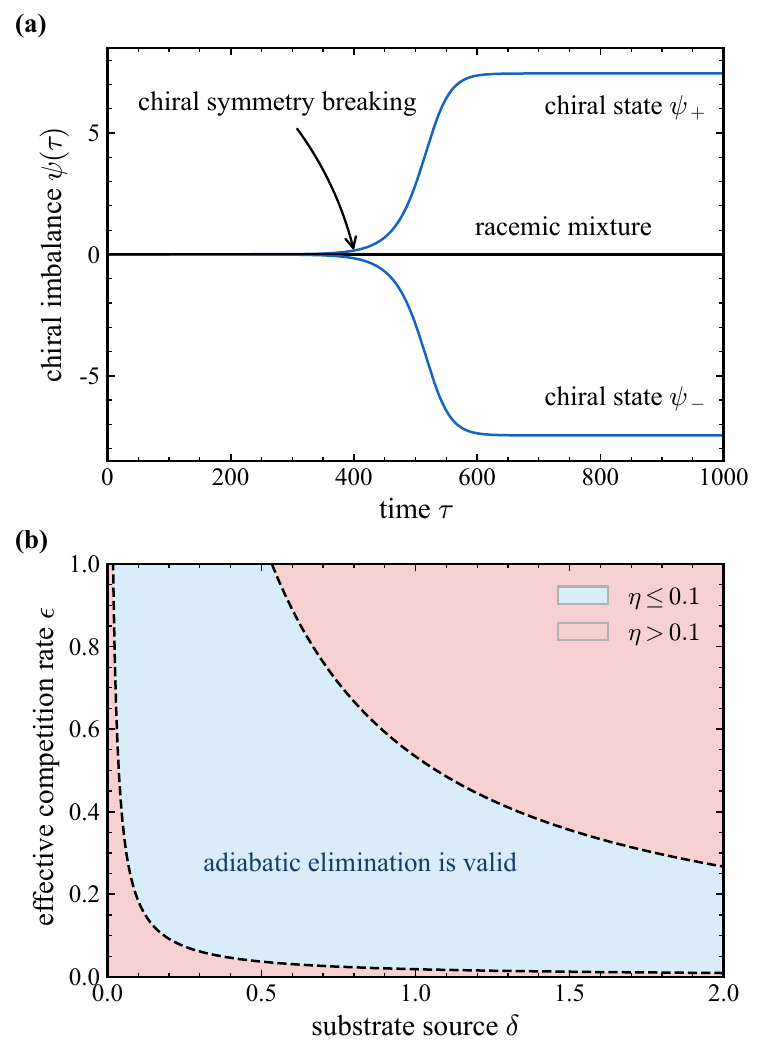}
    \caption{
\textbf{Chiral symmetry breaking in the deterministic Kondenpudi–Nelson model, and validity of adiabatic elimination.}
\textbf{(a)} Deterministic breaking of chiral symmetry induced by an initial (weak) chiral excess (bias). The black line shows the long-time trajectory for symmetric initial conditions, $\psi_0 = 0$. 
The blue lines are the trajectories when the system is initialized with a small enantiomeric excess, i.e., $\psi_0 = \pm 10^{-6}$. 
These weak initial-condition biases are amplified by the nonlinear chemical reactions, breaking chiral symmetry. Moreover, in deterministic simulations, the final state is always dominated by molecules corresponding to the initially favored enantiomer. 
    \textbf{(b)} Parameter space in $\delta$-$\epsilon$ defining the domains where the adiabatic approximation is valid, with the effective competition rate $\epsilon = k_5/k_1$ and the substrate source $\delta = a$.
    In the blue domain, the relaxation timescale of the total concentration variable ($\phi$) is considerably shorter than that of the chiral imbalance variable ($\psi$). The parameters were set to $\omega = 0.01$, $\sigma = 1.0$, and $\gamma = 0.9$.}
    \label{fig:1}
\end{figure}

This bistable behavior can also be directly observed from numerical simulations of Eqs.~(\ref{eq:6}). When the system is initialized exactly at the symmetric fixed point (Eq. (\ref{eq:A1_1})), it remains in that state. However, even a very small initial enantiomeric excess is amplified by the autocatalytic dynamics, driving the system toward one of the asymmetric chiral states, as illustrated in Fig.~\ref{fig:1}(a); the result is the break of chiral symmetry. 

An important point should be emphasized here: within the deterministic description, the initial enantiomeric excess completely determines the final chiral state of the system. In other words, if the initial condition contains an excess of X, the system evolves toward the chiral state dominated by X, whereas an initial excess of Y drives the system toward the opposite chiral state.

\subsection{The Fokker-Planck equation}

Here, our goal is to understand the importance of fluctuations and their effects on the deterministic picture. In principle, this can be achieved through the solution of the Master equation (Eq. (\ref{eq:4})), which provides the probability distribution $P(\mathbf{q},t)$. However, obtaining analytical solutions directly from the Master equation is usually cumbersome due to the high dimensionality of the state space and the presence of nonlinearities. A valuable alternative is to approximate the Master equation by a Fokker–Planck equation and seek solutions for it.

Since the deterministic equations are written in terms of densities, we introduce
$\mathbf r=\mathbf q/N=(x,y)^T$. In this representation, the chemical jumps in state space become
$\boldsymbol{\nu}_i/N$. The transition rates scale as
$W_i(\mathbf q)=Nw_i(\mathbf r)$, the Kramers--Moyal expansion yields a
Fokker-Planck equation:
\begin{equation}\label{eq:7}
\begin{aligned}
\partial_t P(\mathbf{r},t) = - \sum_{\alpha} \partial_{\alpha} [\mathcal{A}_{\alpha}(\mathbf{r})P] + \frac{1}{2} \sum_{\alpha,\beta} \partial^{2}_{\alpha \beta} [\mathcal{B}_{\alpha \beta}(\mathbf{r})P],
\end{aligned}
\end{equation}
where, the chemical drift
\begin{equation}\label{eq:8}
\mathcal{A}_{\alpha} = \sum_{i=1}^{R} S_{\alpha i}w_i(\mathbf{r}),
\end{equation}
and diffusion in chemical state space
\begin{equation}\label{eq:9}
    \mathcal{B}_{\alpha \beta} = \frac{1}{N}\sum_{i=1}^{R} S_{\alpha i}S_{\beta i}w_i(\mathbf{r}),
\end{equation}
with the stoichiometric matrix $\textbf{S}$ defined in Eqs.~(\ref{eq:1}- \ref{eq:2}). The indices $i=1,\ldots,R$ label the reaction channels, while
$\alpha,\beta=1,\ldots,M$ label the chemical species. The quantity
$S_{\alpha i}$ denotes the element of the stoichiometric matrix, Eq. (\ref{eq:2}), associated
with species $\alpha$ and reaction $i$. 

Considering the symmetry of the system it is useful to rewrite Eq.~(\ref{eq:7}) in terms of the total concentration and chiral imbalance variables $(\phi,\psi)$. To this end, we define
\begin{equation*}
    \mathbf z =
\begin{pmatrix}
\phi\\
\psi
\end{pmatrix},
\qquad
\mathbf r =
\begin{pmatrix}
x\\
y
\end{pmatrix},
\qquad
\mathbf z = J\mathbf r,
\end{equation*}
where the mapping $\mathbf{r} \rightarrow \mathbf{z}$ reads
\begin{equation*}
    J=
\begin{pmatrix}
1 & 1\\
1 & -1
\end{pmatrix}.
\end{equation*}

Since this change of variables is linear, the transformed drift vector and
diffusion matrix are given by
\begin{equation}\label{eq:10}
    \tilde{\boldsymbol{\alpha}}(\mathbf z)= J\mathbf {\mathcal{A}}(\mathbf r),
\qquad
\tilde{\boldsymbol{\beta}}(\mathbf z)= J\mathbf {\mathcal{B}}(\mathbf r)J^T.
\end{equation}

We also rescale time $\tau=k_1t$, which leads to $\boldsymbol{\alpha}(\mathbf{z}) = \tilde{\boldsymbol{\alpha}}(\mathbf{z})/k_1$ and  $\boldsymbol{\beta}(\mathbf{z}) = \tilde{\boldsymbol{\beta}}(\mathbf{z})/k_1$. After introducing these transformations and rescaling the parameters, the Fokker-Planck equation becomes
\begin{align}
\nonumber
\partial_{\tau} P(\phi,\psi,\tau) =\;& -\partial_\phi( \alpha_\phi P)-\partial_{\psi} (\alpha_\psi P) 
 + \frac{1}{2} \partial^{2}_{\phi}(\beta_{\phi \phi}P) \\
 \label{eq:11}
&+ \partial^2_{\phi \psi}(\beta_{\phi \psi}P)
+ \frac{1}{2} \partial^{2}_{\psi}(\beta_{\psi \psi}P),
\end{align}
with the following explicit drift and diffusion terms
\begin{equation}
\begin{split}
    \alpha_\phi(\mathbf z) &= 2\delta + (\sigma - \gamma)\phi
    - \frac{\omega+\epsilon}{2}\phi^{2}
    + \frac{\epsilon-\omega}{2}\psi^{2},\\
    \alpha_\psi(\mathbf z) &= \psi((\sigma - \gamma)-\omega \phi),\\
     \beta_{\phi \phi}(\mathbf z) &= N^{-1} (2\delta + (\sigma + \gamma)\phi + \frac{\omega}{2}(\phi^{2}+\psi^{2})+ \epsilon (\phi^{2}-\psi^{2})),\\
     \beta_{\psi \psi}(\mathbf z) &= N^{-1} (2\delta + (\sigma + \gamma)\phi + \frac{\omega}{2}(\phi^{2}+\psi^{2})),\\
     \beta_{\phi \psi}(\mathbf z) &= N^{-1} \psi((\sigma+\gamma) + \omega\phi).
\end{split}
\end{equation}

Equation~(\ref{eq:11}) does not generally admit an analytical solution, because it is a partial differential equation with first and second order derivatives, non-linear, and mixed terms.
We therefore focus on the regime in which the total-density variable $\phi$
relaxes much faster than the chiral imbalance variable $\psi$~\cite{kondepudi1984chiral}. Under this quasi-static approximation, which allows and adiabatic elimination, the dynamics is projected onto the slow manifold
defined by $\partial_t \phi = \alpha_\phi(\phi,\psi)=0.$ The corresponding positive fixed point is
\begin{equation}\label{eq:13}
    \phi^{*}(\psi) = \frac{(\sigma-\gamma) + \sqrt{(\sigma-\gamma)^2+(\omega + \epsilon)[4\delta - (\omega-\epsilon)\psi^2]}}{\omega+\epsilon}.
\end{equation}

To leading order in the time-scale separation, the reduced drift and diffusion
coefficients are obtained by evaluating the corresponding coefficients of the
full Fokker-Planck equation on the slow manifold $\phi=\phi^*(\psi)$. The reduced probability density
$p(\psi,\tau)$ then obeys the effective one-dimensional Fokker--Planck equation, as it is demonstrated in  Appendix~\ref{subsec:AppB}:
\begin{equation}\label{eq:14}
    \partial_{\tau} p(\psi,\tau) = - \partial_{\psi}(a(\psi) p) + \frac{1}{2}\partial_{\psi}^{2}(b(\psi)p),
\end{equation}
where we have the asymptotic drift and diffusivity given as
\begin{equation}\label{eq:15}
    a(\psi) = \alpha_{\psi}|_{\phi=\phi^{*}}~\text{and}~~b(\psi) = \beta_{\psi\psi}|_{\phi=\phi^{*}}.
\end{equation}

The effective one-dimensional Fokker-Planck equation, Eq.~(\ref{eq:14}), admits a stationary solution:
\begin{equation}\label{eq:16}
p_{\mathrm{st}}(\psi) = \frac{\mathcal{N}}{b(\psi)} \exp \left[ \int^{\psi} \frac{2 a(u)}{b(u)} du\right],
\end{equation}
where $\mathcal{N}$ is the normalization constant. Because the $x,y \geq 0$ are positive concentrations, the transformed
variables obey $\phi\geq 0$ and $|\psi|\leq \phi$, see Eq.~(\ref{eq:A1_40}). After the adiabatic
elimination, the physically allowed values of $\psi$ are therefore those
satisfying $|\psi|\leq \phi^*(\psi)$.

\textit{Validity of the adiabatic elimination.}
 A useful measure of the separation between these time scales
is given by the ratio
\begin{equation}\label{eq:17}
    \eta = \frac{|\lambda_\psi|}{|\lambda_{\phi}|},
\end{equation}
where $\lambda_\psi$ and $\lambda_{\phi}$ are the eigenvalues of the
linearized dynamics, evaluated at the chiral state, associated with the chiral imbalance and total-density
directions, respectively. The adiabatic elimination is expected to be valid
when 
\begin{equation*}
    \eta \ll 1.
\end{equation*}

The parameter regimes satisfying this condition are shown in Fig.~\ref{fig:1}(b). In
particular, parameters in the blue region correspond to $\eta<0.1$, for which
the adiabatic elimination provides a very good approximation, as will be
demonstrated below.

\subsection{Stochastic simulations}\label{subsec:StS}

Throughout this work, we performed stochastic numerical simulations of the
reaction network to support our analytical results. 
To this end, we use a Gillespie algorithm~\cite{gillespie1977exact}. 
This method generates numerical realizations of the stochastic process $\mathbf{q}(t)$ by sampling both the waiting times between successive reactions and the reaction channel to occur at each step according to the transition rates.
From an ensemble of such trajectories, we construct an empirical approximation
to the probability distribution $P(\mathbf{q},t)$, which corresponds to the solution of the Master equation (Eq.~(\ref{eq:4})). 
For all cases shown
in this work, statistical averages were computed over 50~000
independent realizations.

\section{Results and Discussion}

\subsection{Effective potential}

We first discuss the stationary probability distribution depicted in Eq.~(\ref{eq:16}). 
This distribution can be written in terms of an effective potential, $U(\psi)$:
\begin{equation}
    p_{\mathrm{st}}(\psi) = \mathcal{N} e^{U(\psi)},
\end{equation}
with the effective potential 
\begin{equation}\label{eq:19}
    U(\psi) = - \int^{\psi}du~\frac{2a(u)}{b(u)} + \ln b(u).
\end{equation}

\begin{figure}[t]
    \centering
    \includegraphics[width=0.7\linewidth]{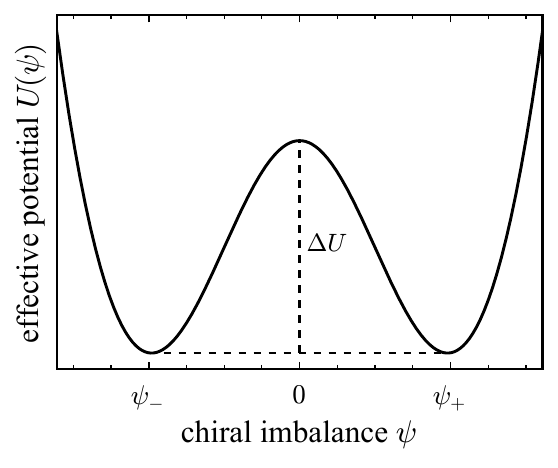}
    \caption{\textbf{Effective double-well potential for chiral symmetry breaking}. 
    Stochastic chiral symmetry breaking in a racemic state is analogous to a Brownian particle moving in a potential well. 
    The two minima are located in the chiral states $\psi_-$ and $\psi_+$, and the maximum is located at the racemic state, i.e., $\psi = 0$. The potential barrier height is $\Delta U$. 
     }
    \label{fig:2}
\end{figure}

As expected from the intrinsic bimodal nature of this problem, this potential is a symmetric double-well potential, as shown in Fig.~\ref{fig:2}. The extrema are located at the fixed points of $\psi$, i.e., $\psi^*$ which belongs to the set $ \{ \psi_-,~0,~ \psi_+\}$, with two minima corresponding to the chiral states, confirming the stability of these fixed points, and a maximum at the saddle-node, i.e., racemic state, fixed point. 
Therefore, the state of the system can be understood as a Brownian particle moving in a double-well potential~\cite{gardiner2009stochastic}. This interpretation can guide us in addressing chiral symmetry breaking.

One useful quantity that can be extracted from this potential is the height of the potential barrier, $\Delta U(\psi)$, which is given by
\begin{equation}\label{eq:20}
    \Delta U(\psi) = \int_{0}^{\psi_+} du~ \frac{2a(u)}{b(u)} + \ln  \left[ \frac{b(0)}{b(\psi_+)} \right],
\end{equation}

where $\psi_+$ is the positive chiral fixed point. This quantity provides information about the depth of the basin of attraction associated with the chiral states. Larger values of $\Delta U$ correspond to more stable states, making transitions between opposite chiral states less likely. In contrast, smaller values of $\Delta U$ indicate that the system can transition more easily from one chiral state to the other. Note that, by symmetry, the potential barrier is the same for both chiral states.

\subsection{Bimodal distributions}

\begin{figure}[t]
    \centering
    \includegraphics[width=1.0\linewidth]{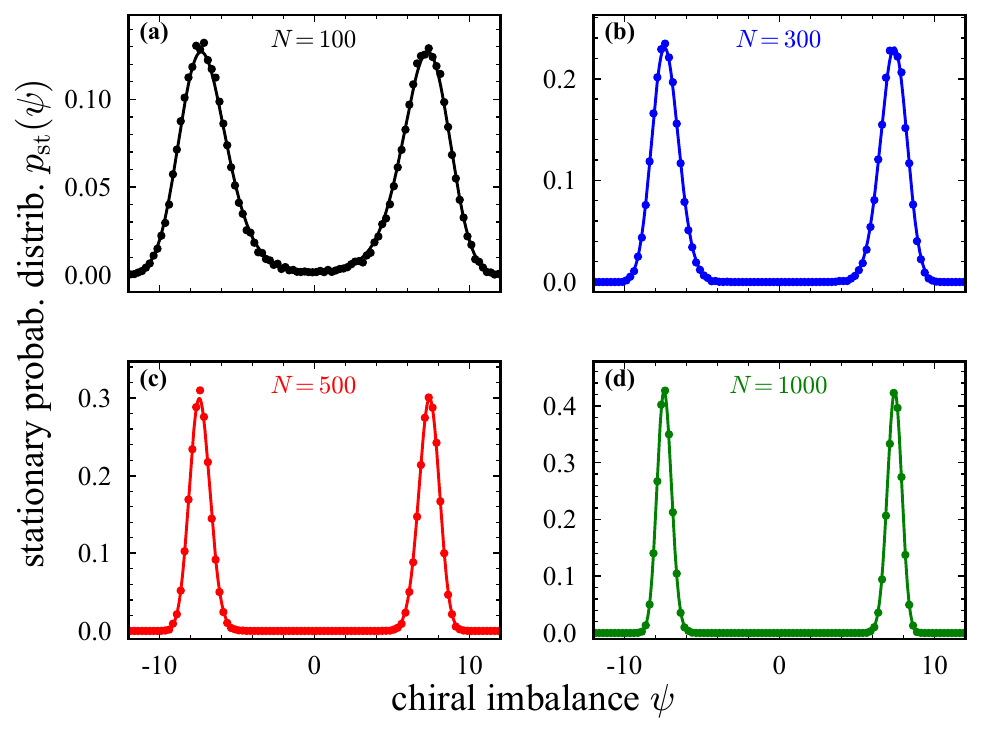}
    \caption{ \textbf{Steady state bimodal distributions of the stochastic Kondepudi-Nelson model}.
    The solid lines come from the stationary solution of the Fokker-Planck equation (Eq.~\eqref{eq:16}), while the dots are the ensemble average of the numerical stochastic trajectories, as described in Sec.~\ref{subsec:StS}. 
    The peaks of the bimodal distributions are located at the minima of the effective potential; see Fig.~\ref{fig:2}.
    Increasing the number of molecules $N$ narrows the width around each distribution's peaks corresponding to the deterministic solutions, where fluctuations can be neglected.}
    \label{fig:3}
\end{figure}

For an unbiased system, the distribution of the chiral imbalance is expected to be
a symmetric bimodal distribution. This behavior is confirmed in Fig.~\ref{fig:3}(a-d), where
the stationary distribution is plotted for different numbers of particles in a
region of parameter space where the adiabatic elimination is valid. In the same
figure, we also show the results obtained from stochastic numerical simulations
of the full model. These simulations were performed using the racemic state as
the initial condition, allowing the final state to be selected by intrinsic
fluctuations. We find excellent agreement between the analytical results (solid lines) and the
numerical simulations (symbols), confirming the validity of the time-scale separation.

From Fig.~\ref{fig:3}, we see that the maxima of the bimodal distributions are
located at, or very close to, the asymmetric fixed points, $ \psi_-$ and $\psi_+$. These points
also correspond to the two minima of the effective potential shown in Fig.~\ref{fig:2}.
Furthermore, the stationary distributions become narrower as the number of
particles increases. This dependence on the number of particles can be understood
by approximating the dynamics near the peaks of the distribution. At a stable
fixed point $\psi^*$, the effective drift satisfies $ a(\psi^*) = 0 $ and $a'(\psi^*)<0$.

Therefore, near the fixed point $\psi^*$, we may approximate the drift as
\begin{equation}
    a(\psi) \simeq a'(\psi^*)(\psi-\psi^*),
\end{equation}
and the diffusion coefficient as a constant,
\begin{equation}
    b(\psi)\simeq b(\psi^*)>0.
\end{equation}

Under these approximations, the steady-state solution of Eq.~(\ref{eq:14})
is locally Gaussian,
\begin{equation}
    p_{\mathrm{st}}(\psi)
    \propto
    \exp\left[
    -\frac{(\psi-\psi^*)^2}{2\sigma^2}
    \right],
\end{equation}
with mean $\psi^*$ and variance
\begin{equation}
    \sigma^2
    =
    -\frac{b(\psi^*)}{2a'(\psi^*)}.
\end{equation}

Since $b(\psi^*)\propto N^{-1}$, the variance decreases as the number of
particles increases. This explains the narrowing of the stationary distributions
observed for larger values of $N$. In the limit of very large $N$, the variance
approaches zero and the probability distribution converges to a pair of Dirac
delta functions centered at the stable deterministic fixed points. Therefore, in
finite systems, drift and diffusion compete. In contrast, the deterministic dynamics
selects the stable fixed points, intrinsic fluctuations broaden the stationary
distribution around them and can induce transitions between the two asymmetric
states.

\subsection{Initial enantiomeric excess vs. fluctuations}

As discussed in Sec.~\ref{subsec:IIC}, the initial enantiomeric excess determines the configuration of the final state in the thermodynamic limit, where chiral symmetry breaking is driven predominantly by deterministic drift. However, because the system considered here is intrinsically stochastic, the competition between drift and diffusion may modify the deterministic prediction. We therefore ask how the initial enantiomeric excess affects the probability of selecting a given final homochiral state in the presence of fluctuations.

This problem has been addressed numerically before~\cite{silva2017stochastic,hochberg2023chiral}. It was found that, in finite populations, the initially more abundant enantiomer is not always the one selected in the final homochiral state . Here, we seek an analytical understanding of this effect.

We consider the following question: what is the probability that the system first reaches the chiral state $\psi_+$, given an initial condition $\psi(0)=\psi_0$? By symmetry, the corresponding probability of first reaching $\psi_-$ follows analogously. This can be formulated as a first-passage problem. Let $\Pi(\psi_0)$ be the probability that, starting from $\psi(0)=\psi_0 \in (\psi_-,\psi_+)$, the process reaches the boundary $\psi_+$ before $\psi_-$. The corresponding backward equation is
\begin{equation}
    a(\psi_0) \Pi'(\psi_0) + \frac{1}{2}b(\psi_0)\Pi''(\psi_0) = 0.
\end{equation}

A complete derivation of this equation is given in Appendix~\ref{subsec:AppC}.  It is noteworthy that in the backward formulation the the probability $\Pi(\psi_0)$ is a function of the initial condition. The construction of the first-passage problem imposes the boundary conditions
$\Pi(\psi_-)=0$ and $\Pi(\psi_+)=1$. The solution is~\cite{gardiner2009stochastic}
\begin{equation}\label{eq:24}
    \Pi(\psi_0) =
    \frac{\int_{\psi_-}^{\psi_0} \exp[\Phi(s)]\,ds}
    {\int_{\psi_-}^{\psi_+} \exp[\Phi(s)]\,ds},
\end{equation}
where $\Pi(\psi_0)$ is also known as the effective splitting probability and 
\begin{equation}\label{eq:25}
    \Phi(\psi_0) =
    - \int^{\psi_0} du \, \frac{2a(u)}{b(u)}.
\end{equation}

\begin{figure}[tb]
    \centering
    \includegraphics[width=0.92\linewidth]{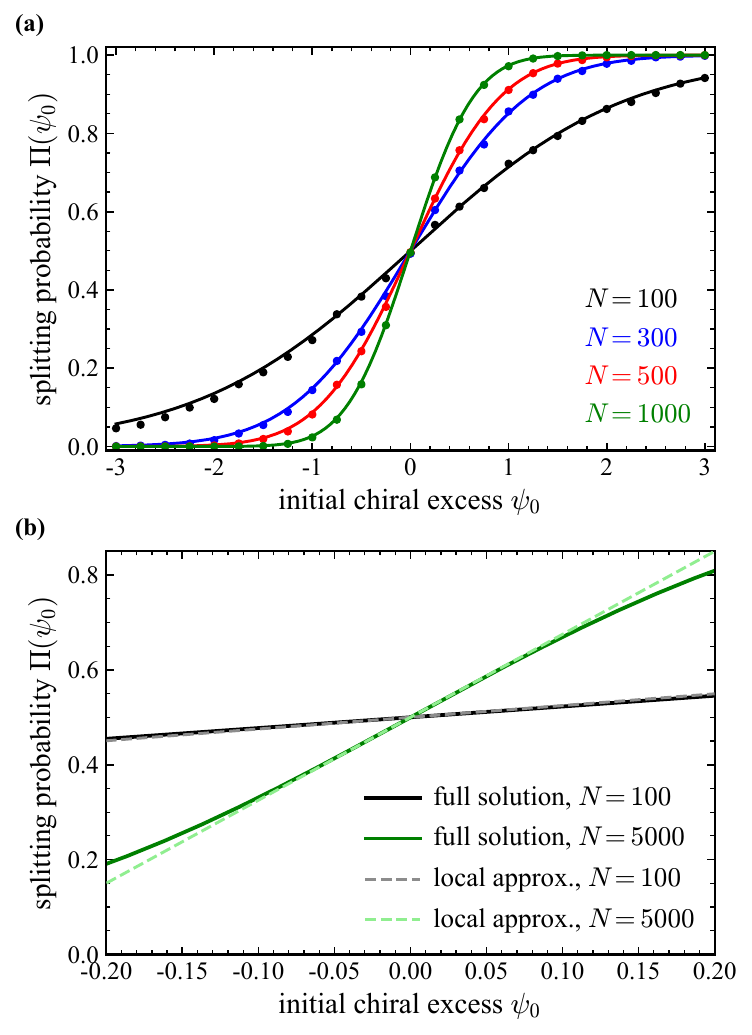}
 \caption{
 \textbf{Competition between initial enantiomeric excess and fluctuations.}
 \textbf{(a)} Splitting probability for different population sizes $N$. The solid lines were obtained analytically (Eq.~\eqref{eq:24}), while the dots represent ensemble averages over trajectories obtained from stochastic simulations, as described in Sec.~\ref{subsec:StS}. 
 Finite-population fluctuations compete with the initial enantiomeric excess $\psi_0$, probabilistically favoring the initially more abundant enantiomer in the final state. 
 This competition is especially evident for small populations, and the transition sharpens as the number of molecules increases. 
 In the thermodynamic limit (large $N$), any finite initial excess strictly favors the corresponding enantiomer, recovering the deterministic predictions.
 \textbf{(b)} Approximation of the splitting probability in the vicinity of $\psi_0 = 0$ (Eq.~\eqref{eq:26}). This approximation enables the identification of two limiting regimes: (1) the small-population regime (low $N$), in which fluctuations dominate the selection process and the probability of reaching either chiral state is determined by the initial state within the interval $(\psi_-,\psi_+)$; and (2) the large-population regime (large value of $N$), in which the slope of the probability of reaching the chiral states is controlled by the ratio between the initial enantiomeric excess and the typical fluctuation amplitude, $\psi_0/\delta\psi$. }
    \label{fig:4}
\end{figure}

The analytical solution, Eq.~\eqref{eq:24}, is shown in Fig.~\ref{fig:4}(a) together with numerical simulations. 
The curve has a sigmoidal shape and satisfies $\Pi(0)=1/2$, indicating that, in the absence of an initial bias, both homochiral configurations are equally likely. As $\psi_0$ approaches $\psi_+$, the probability $\Pi(\psi_0)$ approaches $1$, as required by the boundary condition. Conversely, as $\psi_0$ approaches $\psi_-$, the probability tends to $0$.

These results show that finite-population fluctuations can strongly affect final-state selection. The initial enantiomeric excess and intrinsic fluctuations compete. Although increasing $\psi_0$ makes selection of the $\psi_+$ state more likely, it does not completely exclude the possibility that the final state is dominated by the initially less abundant enantiomer. However, such events become less likely as the initial bias increases. Fig.~\ref{fig:4}(a) also shows that this probability decreases for larger values of $N$, as the transition in $\Pi(\psi_0)$ becomes sharper around $\psi_0=0$.

Further insight can be obtained by expanding Eq.~\ref{eq:24} near $\psi_0=0$. In this region, the solution is approximately linear. As shown in Appendix \ref{subsec:AppD}, a first-order Taylor expansion gives
\begin{equation}\label{eq:26}
    \Pi(\psi_0) \simeq
    \frac{1}{2}
    +
    \frac{\sqrt{\kappa}}
    {\sqrt{\pi} \erf (\psi_+\sqrt{\kappa})}
    \psi_0 .
\end{equation}
Here, $\kappa=a'(0)/b(0)$ and $\erf(x)$ denotes the error function. Since $b \propto N^{-1}$, we can write $\kappa=cN$, where $c \in \mathbb{R}^+$. Figure~\ref{fig:4}(b) compares this approximation with the full solution. The approximation agrees well close to $\psi_0=0$, while deviations appear as the initial enantiomeric excess increases.

Equation \eqref{eq:26} can be analyzed in two limiting regimes:
(i) When $cN \ll 1$, or equivalently $\psi_+\sqrt{cN} \ll 1$, we use
$\erf(x)\simeq 2x/\sqrt{\pi}$ and obtain
\begin{equation}
     \pi(\psi_0) \simeq
     \frac{1}{2}
     \left(
     1+\frac{\psi_0}{\psi_+}
     \right).
\end{equation}

This regime can occur when the deterministic drift is weak near $\psi=0$, or when the population size is sufficiently small. In this limit, drift is negligible compared with diffusion in determining the splitting probability. As a consequence, the probability of reaching either absorbing state depends only on the relative position of the initial condition within the interval $(\psi_-,\psi_+)$, and not explicitly on the population size $N$. This does not mean that fluctuations are irrelevant; rather, fluctuations dominate the selection process. The amplitude of fluctuations may still affect the time required to reach a homochiral state, but it cancels out of the splitting probability in this drift-dominated approximation.

The other limiting regimes (ii) corresponds to  $cN \gg 1$, more precisely when $\psi_+\sqrt{cN} \gg 1$, we have $\erf(\psi_+\sqrt{cN})\approx 1$. Equation~\ref{eq:26} then becomes
\begin{equation}
     \pi(\psi_0) \simeq
     \frac{1}{2}
     +
     \sqrt{\frac{cN}{\pi}}\,\psi_0 .
\end{equation}

In this regime, the slope of the splitting probability near $\psi_0=0$ scales as $\sqrt{N}$. Since the relative amplitude of fluctuations scales as
$\delta\psi \sim N^{-1/2}$, this result suggests that the relevant control parameter is the ratio between the initial enantiomeric excess and the typical fluctuation amplitude, i.e., $\psi_0/\delta\psi$. Thus, for larger populations, fluctuations are weaker and the probability curve becomes steeper near $\psi_0=0$. Conversely, for smaller populations, fluctuations are stronger and the dependence of the final state on the initial enantiomeric excess becomes smoother, as observed in Fig.~\ref{fig:4}(a).

\subsection{Mean time to reach the chiral state}

After examining how the initial chiral excess affects the definition of the final chiral state, it is also relevant to evaluate how long the system takes to reach one of the chiral states (\(\psi_+\) or \(\psi_-\)) for different initial conditions. 
More precisely, we ask: what is the mean time required for the system to reach one of the chiral states as a function of the initial condition? This question can be addressed in a manner similar to that used in the previous subsection.

Let \(T(\psi_0)\) denote the mean time required for the system to reach one of the chiral states. This quantity can be obtained from the following backward equation:

\begin{equation}\label{eq:29}
    a(\psi_0)T'(\psi_0) + \frac{1}{2}b(\psi_0)T''(\psi_0) = -1,
\end{equation}

whose derivation is presented in Appendix \ref{subsec:AppE}. Solving this equation subject to the absorbing boundary conditions \(T(\psi_-)=T(\psi_+)=0\) yields

\begin{equation}\label{eq:30}
\begin{split}
    T(\psi_0) =
    C\int_{\psi_-}^{\psi_0}\exp[-\Phi(s)]\,ds
    \\
    -\int_{\psi_-}^{\psi_0}\exp[-\Phi(s)]
    \int_{\psi_-}^{s}
    \frac{2\exp[\Phi(u)]}{b(u)}
    \,du\,ds ,
\end{split}
\end{equation}

with the normalization 

\begin{equation}
    C=
    \frac{
    \int_{\psi_-}^{\psi_+}
    \exp[-\Phi(s)]
    \int_{\psi_-}^{s}
    \frac{2\exp[\Phi(u)]}{b(u)}
    \,du\,ds
    }
    {
    \int_{\psi_-}^{\psi_+}
    \exp[-\Phi(s)]\,ds
    }.
\end{equation}

\begin{figure}[tb]
    \centering
    \includegraphics[width=0.9\linewidth]{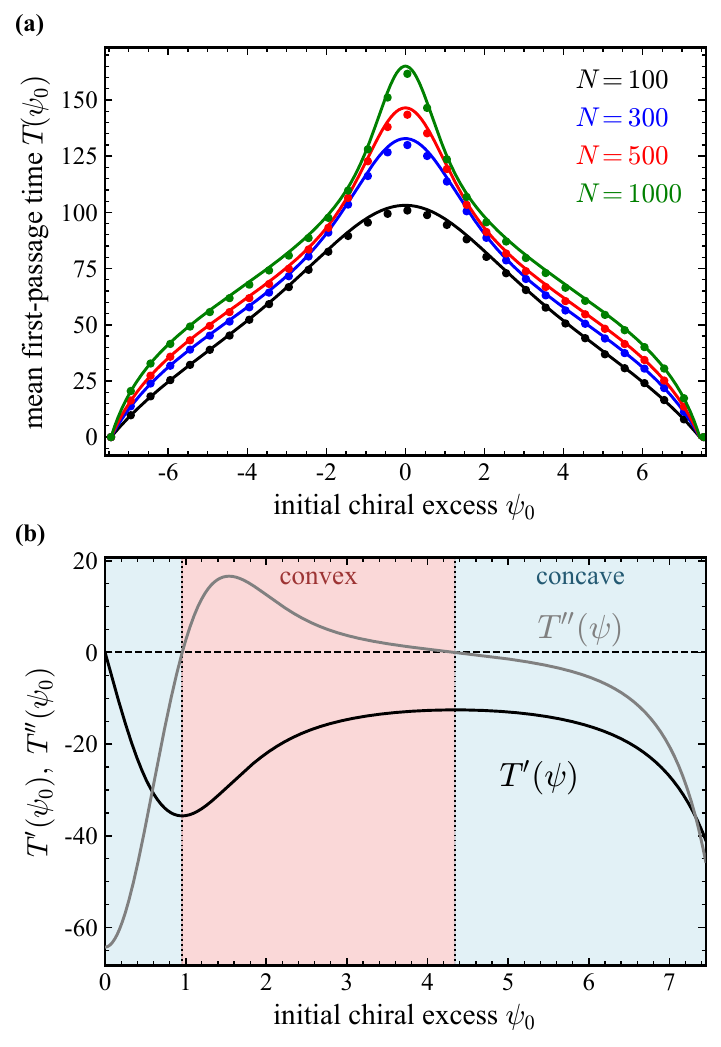}
    \caption{\textbf{Chiral symmetry braking time scales for different initial conditions and population sizes.}
    \textbf{(a)} Mean first-passage time to reach the chiral states. The solid lines are analytical (Eq.~\eqref{eq:30}), while the dots represent ensemble averages over trajectories from stochastic simulations, as described in Sec.~\ref{subsec:StS}. Larger populations $N$ require longer times to reach the chiral states. In all cases, the mean first-passage time is longest for the symmetric initial state and decreases in a nontrivial manner as the initial enantiomeric bias approaches either chiral state. The cusp structure of the mean first-passage-time function reflects the abrupt change in the direction of the deterministic dynamics, which is associated with the symmetry of the system.
\textbf{(b)} First and second derivatives, $T'(\psi_0)$ and $T''(\psi_0)$, of the mean first-passage-time solution. The function $T(\psi_0)$ exhibits nontrivial behavior in both intervals, $(\psi_-,0)$ and $(0,\psi_+)$. For small enantiomeric excesses, the mean first-passage time decreases rapidly. As $|\psi_0|$ increases, however, this decrease becomes less pronounced, as indicated by a change in the curvature of the function from concave to convex. A further increase in $|\psi_0|$ leads to another transition in the curvature of $T(\psi_0)$, accompanied by a more pronounced decrease in the mean first-passage time as the initial condition approaches the chiral states.}
    \label{fig:5}
\end{figure}

In the equations above, \(\Phi(\psi)\) is given by Eq.~(\ref{eq:25}). The analytical and numerical mean first-passage times to reach one of the chiral states are shown in Fig.~\ref{fig:5}(a). As expected, larger population sizes require longer times to reach the absorbing boundaries. Furthermore, as \(N\) increases, \(T(\psi_0)\) develops a pronounced cusp around \(\psi_0=0\). This structure reflects the abrupt change in the direction of the deterministic dynamics, i.e., depending on the sign of the initial bias, the system evolves toward either \(\psi_-\) or \(\psi_+\). 
As the thermodynamic limit is approached, the cusp becomes increasingly sharp, indicating the emergence of a non-differentiable point at \(\psi_0=0\). This behavior is consistent with the deterministic description, in which the racemic state remains indefinitely occupied (or at very larger time scales in that point) when the system is initialized exactly at \(\psi_0=0\).

Although the cusp structure is most evident close to the racemic state, Fig.~\ref{fig:5}(a) also shows that the curvature of \(T(\psi_0)\) changes as the initial enantiomeric excess increases. 
This feature is highlighted in Fig.~\ref{fig:5}(b), where \(T'(\psi_0)\) and \(T''(\psi_0)\) are shown for \(N=500\). 
Very close to \(\psi_0=0\), \(T(\psi_0)\) is concave, and small differences in the initial bias lead to significant changes in the mean first-passage time, as reflected by the steep slope of the curve. 
As \(\psi_0\) increases, an inflection point is reached and the curvature changes sign. 
Beyond this point, and up to the next change in curvature, the derivative \(T'(\psi_0)\) remains relatively small, indicating a regime in which substantial variations in the initial condition produce only modest changes in the mean first-passage time.
For larger values of the initial enantiomeric excess, a second inflection point is observed, after which \(T(\psi_0)\) becomes concave again. 
Finally, near the absorbing boundaries, the magnitude of the slope increases sharply, reflecting the strong dependence of the mean first-passage time on the initial condition in this region.

\begin{figure}[t]
    \centering
    \includegraphics[width=0.95\linewidth]{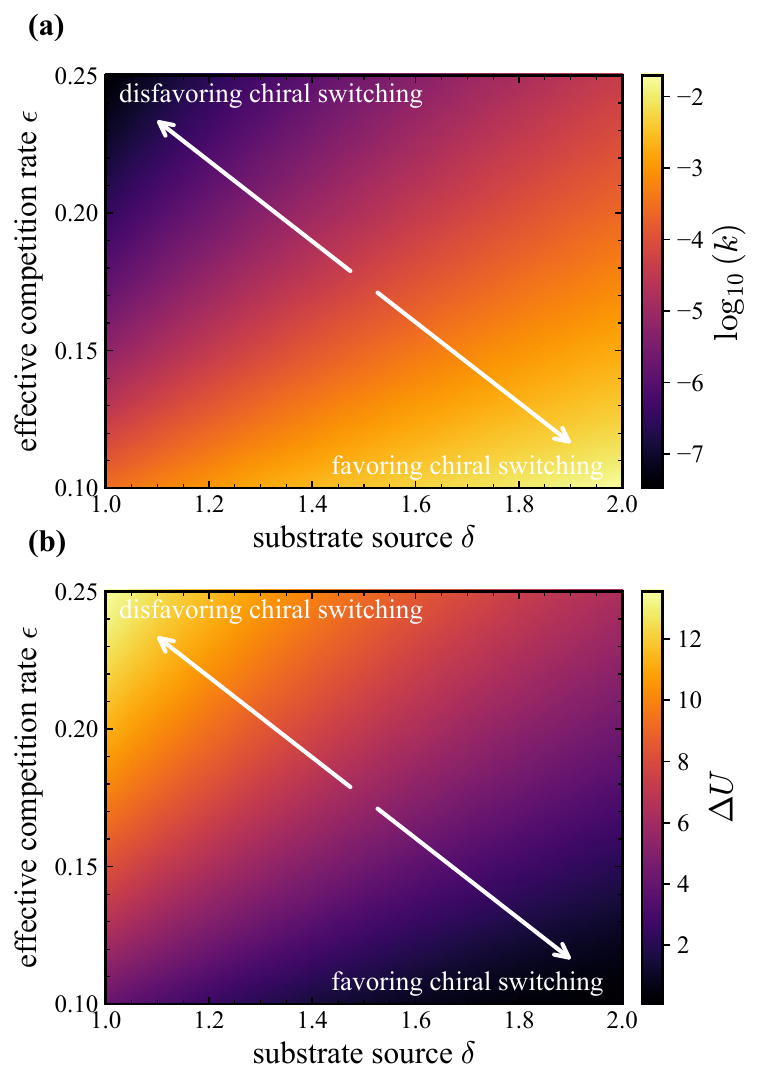}
    \caption{\textbf{Conditions for chiral switching.}
    \textbf{(a)} Escape rate $k$ for $\omega = 0.01$, $\gamma = 0.9$, $\sigma = 1.0$, and different values of $\delta$ and $\epsilon$. Interconversion between the two chiral states is favored when the production rate of enantiomers from chiral substrates, represented by $\delta$, is high and the effective competition rate, represented by $\epsilon$, is low. Under these conditions, the escape rate is small, allowing fluctuations to induce transitions between the two chiral states.
\textbf{(b)} Height of the effective potential barrier $\Delta U$ under the same conditions as in (a). Interconversion between the two chiral states is favored when the effective potential barrier separating them is small. Lower barriers are obtained under the same conditions for which small escape rates are observed in (a), demonstrating the correspondence between the effective potential landscape and the fluctuation-induced transitions between chiral states.}
    \label{fig:6}
\end{figure}

\subsection{Interconversion of chiral configurations}

As suggested in the previous sections, the double-well potential indicates the possibility of interconversion between chiral configurations in the presence of fluctuations. Therefore, one may ask: what is the rate at which the system escapes from a chiral state and transitions to the opposite configuration? We address this question by evaluating the mean first-passage time for a system initially located at $\psi_0=\psi_-$ to cross the potential barrier at $\psi=0$. The solution can be obtained from Eq.~(\ref{eq:29}) using the boundary conditions $T(0)=0$ and $T'(\psi_-)=0$, yielding
\begin{equation}\label{eq:32}
    T(\psi_0) = 2 \int_{\psi}^{0} ds\,\exp[\Phi(s)]
    \int_{\psi_-}^{s} du\, \frac{\exp[-\Phi(u)]}{b(u)},
\end{equation}
where $\Phi(\psi)$ is given by Eq.~(\ref{eq:25}). From this expression, we calculate the escape-rate constant $k$ as
\begin{equation*}
    k=T(\psi_+)^{-1}.
\end{equation*}

The results are shown in Fig.~\ref{fig:6}. Figure~\ref{fig:6}(a) presents the escape-rate constant $k$ for different values of $\epsilon$ and $\delta$. 
One can observe that smaller values of $\epsilon$ and larger values of $\delta$ lead to larger escape rates. 
Chemically, this indicates that transitions between chiral states are favored in systems where the conversion of substrates into enantiomers is fast and the competitive inhibition process is relatively weak. 
The opposite behavior is observed when competition is strong and the conversion process is slow. 
In such cases, transitions between chiral states become extremely unlikely.

Additional insight can be obtained by deriving an approximate expression for Eq.~(\ref{eq:32}). 
In the weak-noise regime, corresponding to a sufficiently large potential barrier ($\Delta U \gg 1$), the exponential terms become strongly localized around the extrema of $\Phi(\psi)$. 
In this limit, the inner integral is dominated by the vicinity of the barrier top at $\psi=0$, whereas the outer integral is dominated by the vicinity of the stable fixed point $\psi_-$. 
Expanding $\Phi(\psi)$ to second order around these points and applying saddle-point method (see Appendix~\ref{subsec:AppF}), we obtain
\begin{equation}\label{eq:33}
   k
\simeq
\frac{b(0)\sqrt{U''(\psi_-)|U''(0)|}}{\pi}
\exp[-\Delta U].
\end{equation}

Equation~(\ref{eq:33}) shows that the escape rate depends exponentially on the effective potential barrier, $k \propto e^{-\Delta U}$. Consequently, even modest variations in $\Delta U$ can produce orders-of-magnitude changes in the switching rate. 
Therefore, a practical way to understand the interconversion dynamics is through an analysis of the effective potential; 
see Eq.~(\ref{eq:20}). 
In Fig.~\ref{fig:6}(b), we show the values of $\Delta U$ for different values of $\epsilon$ and $\delta$. 
The same qualitative trends observed for the escape rates are recovered here. 
In particular, regions with lower potential barriers correspond to larger escape rates, indicating that it is easier for the system, viewed as an effective Brownian particle, to overcome the barrier separating the two chiral configurations.
Consequently, transitions between chiral states become more frequent as the barrier height decreases.

\begin{figure}[t]
    \centering
    \includegraphics[width=1.0\linewidth]{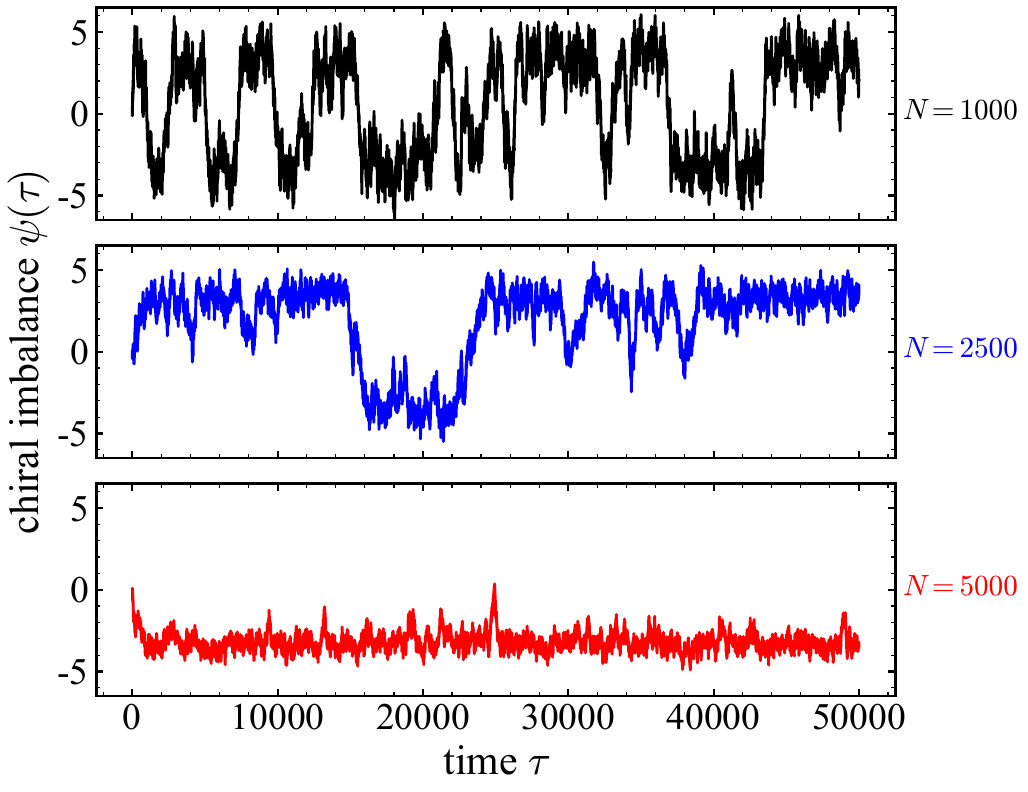}
    \caption{\textbf{Numerical simulations of the Kondepudi-Nelson model in conditions favoring chiral switching}.
    Stochastic trajectories of the chiral imbalance for systems with three different population sizes: $N = 1000$, $N = 2500$, and $N = 5000$. Under the selected conditions, the chiral states are located at $\psi_{\pm} = \pm 3.3$. Transitions between the chiral states are more frequent in systems exhibiting larger fluctuations, i.e., systems with smaller populations. As the number of particles increases, these transitions become increasingly rare and are no longer observed over practical timescales. This behavior suggests that, in the thermodynamic limit, such transitions are suppressed, highlighting the intrinsically stochastic nature of this phenomenon. Parameters selected: $\omega = 0.01$, $\gamma = 0.9$, $\sigma = 1.0$, $\epsilon = 0.1$, and $\delta = 2.0$. }
    \label{fig:7}
\end{figure}

To validate these predictions, we selected parameter values for which the effective potential barrier is relatively small and performed stochastic simulations. 
The results are shown in Fig.~\ref{fig:7} for different system sizes. 
For small values of $N$, the system frequently switches between the two chiral configurations, often undergoing multiple transitions during a single realization. 
As the number of particles increases, however, these switching events become progressively rarer and eventually disappear within the simulation time window. 
This behavior indicates that finite-size effects play a crucial role in enabling transitions between chiral states.

\section{Conclusion}

Here, we have shown that finite-population fluctuations strongly influence chiral selection in systems in which symmetry breaking is driven by a deterministic instability. Although our results were obtained for a specific reaction scheme, the Kondepudi–Nelson model, the underlying framework is expected to apply more broadly to chemical systems governed by Frank-type autocatalysis and competition. Three main conclusions emerge from our analysis.

(i) Chiral selection is governed by a competition between the initial enantiomeric excess and finite-population fluctuations. Consequently, there is a nonzero probability that the system selects the enantiomer that is initially less abundant. In the diffusion-dominated regime at small system size, this probability depends only on the relative position of the initial state between the two chiral states. For larger systems, selection is instead controlled by the ratio of the initial enantiomeric excess to the characteristic fluctuation amplitude, $\psi_0/\delta\psi$. This result provides a quantitative connection between initial bias, system size, and selection probability. It may therefore be relevant to experiments in which chirality is induced by an initial enantiomeric bias~\cite{soai1995asymmetric,sato2003amplification,soai2014asymmetric}, as it allows the bias required to obtain a given selection probability to be estimated.

(ii) The characteristic time for chiral selection increases with system size and decreases with increasing initial enantiomeric bias. Thus, finite-population fluctuations not only affect which chiral state is selected but also accelerate the selection process. This timescale may be important when considering prebiotic scenarios based on nonequilibrium reaction networks: if symmetry breaking requires sustained nonequilibrium fluxes, the physical environment must persist for at least the characteristic selection time. Finite-population fluctuations can relax this temporal constraint by reducing the time required to reach a chiral state.

(iii) When the production rate of enantiomers from achiral substrates is sufficiently high relative to the competition rate, finite-population fluctuations can induce switching between the two chiral states. The first state selected by the system therefore need not remain stable over long times. The switching rate decreases strongly with system size because the effective potential barrier scales extensively, $\Delta U\propto N$, as follows from Eq.~(\ref{eq:20}). Consequently, switching becomes increasingly suppressed as the thermodynamic limit is approached.

Noise-induced switching between chiral states has also been observed in molecular-dynamics simulations of the chiral tetramer model~\cite{piaggi2023critical}. There, the free-energy barrier scales as $\Delta F\propto N^{2/3}$, consistent with a surface-dominated process involving spatially structured chiral domains. By contrast, our spatially homogeneous description gives $\Delta U\propto N$, corresponding to a collective fluctuation of the system. Despite this difference, both descriptions predict a barrier that diverges with system size and therefore identify chiral switching as a finite-size phenomenon. Our results further show that such switching can arise from the stochastic dynamics of a coarse-grained chemical reaction network, without requiring explicit spatial interfaces or molecular interactions.

Taken together, these results show that finite-population fluctuations play two distinct roles in Frank-type chiral symmetry breaking: they influence the probability and timescale of the initial chiral selection and, under appropriate conditions, can destabilize the selected state through stochastic switching. These effects may be relevant to prebiotic environments in which the effective reacting populations were sufficiently small for intrinsic fluctuations to compete with deterministic dynamics. At the same time, our analysis identifies an important constraint on such scenarios: chiral selection must occur within the lifetime of the nonequilibrium conditions that sustain the symmetry-breaking dynamics.

The present results are specific to systems in which chiral symmetry breaking arises from autocatalysis and competition in a well-mixed reaction network. Other physicochemical mechanisms may generate chiral asymmetry under different conditions. For example, recent experiments have shown that polymers undergoing liquid–liquid phase separation and forming liquid-crystalline states can spontaneously develop global chirality~\cite{xu2025ubiquitous}. Extending stochastic approaches to spatially resolved and phase-separating systems would help establish which aspects of finite-population chiral selection are universal and which depend on the underlying symmetry-breaking mechanism.

\begin{acknowledgements} 
We thank Dr.\,Abhinav Sharma for insightful discussions.
L.\,Silva-Dias acknowledges the CAPES-Humboldt Research Fellowship Program, a joint program of the Coordenação de Aperfeiçoamento de Pessoal de Nível Superior (CAPES Finance Code 001) and the Alexander von Humboldt Foundation, for financial support.
\end{acknowledgements}

\section*{Appendix}

\appendix
\renewcommand{\thesection}{\Alph{section}}
\renewcommand{\theequation}{A\arabic{equation}}

\subsection{Fixed points and linear stability analysis}\label{subsec:AppA}

The fixed points, ($\phi^{*}, \psi^{*}$), from the set of Eqs. (\ref{eq:6}) are determined when $\dot{\phi(\tau)} = \dot{\psi(\tau)} = 0$. There are four possible ones, two symmetric fixed points,
\begin{equation}\label{eq:A1_1}
    (\psi^*, \phi^*) =
\left(
0,\;
\frac{(\sigma-\gamma) \pm \sqrt{(\sigma-\gamma)^2 + 4(\omega+\epsilon)\delta}}{\omega+\epsilon}
\right),
\end{equation}
and two asymmetric fixed points,
\begin{equation}\label{eq:A1_2}
    (\psi^*, \phi^*) =
\left(
\pm \sqrt{\frac{(\sigma-\gamma)^2}{\omega^2} + \frac{4\delta}{\omega-\epsilon}},
\;
\frac{\sigma-\gamma}{\omega}
\right).
\end{equation}

One can easily verify that, depending on the choice of parameters, the fixed points may become either complex or negative. However, here we are interested in physically meaningful situations in which there exists one symmetric fixed point and two asymmetric fixed points. Therefore, we must determine the conditions under which such states exist.

\subsubsection{Existence and positivity conditions}

Considering the physical nature of the system, we assume throughout this work that all parameters are real and positive.
From Eq.~(\ref{eq:A1_2}), the existence of fixed points requires that
\begin{equation}\label{eq:A1_3}
    \omega \neq 0
    \qquad \text{and} \qquad
    \omega \neq \epsilon.
\end{equation}

Let us first analyze the existence of real and positive fixed points for the variable $\phi$. For the symmetric solution we have two fixed points which we address as $\psi_-$ and $\phi_+$. 
In this case we can verify that $\phi_-$ is always  negative since all parameters are assumed to be positive. However, 
\begin{equation}
    \phi_+ =
    \frac{(\sigma-\gamma)
    + \sqrt{(\sigma-\gamma)^2 + 4(\omega+\epsilon)\delta}}
    {\omega+\epsilon},
\end{equation}
can take positive values if the following condition holds
$\sigma > \gamma$.
Under the same condition, the asymmetric fixed points also possess positive values of $\phi^*$.

Let us now analyze the variable $\psi$. 
Unlike $\phi$, the chiral imbalance may assume either positive or negative values; however, it must remain real. 
For the symmetric fixed point, we always have $\psi^*=0$. 
For the asymmetric fixed points, it is required that
\begin{equation}
    \frac{(\sigma-\gamma)^2}{\omega^2}
    +
    \frac{4\delta}{\omega-\epsilon}
    >0.
\end{equation}
This condition ensures the existence of real asymmetric fixed points.

We must additionally guarantee that the original concentrations satisfy $x^*,y^*>0$. Using the transformation
\begin{equation}
x^* = \frac{\phi^*+\psi^*}{2},
\qquad
y^* = \frac{\phi^*-\psi^*}{2},
\end{equation}
this condition above is equivalent to
\begin{equation}\label{eq:A1_40}
\phi^*>0
\qquad \text{and} \qquad
|\psi^*|<\phi^*.
\end{equation}

Combining the expressions for $\phi^*$ and $\psi^*$, we obtain
\begin{equation}
    (\psi^*)^2 - (\phi^*)^2
    =
    \frac{4\delta}{\omega-\epsilon}
    <0.
\end{equation}
Since $\delta>0$, this condition implies
$\epsilon > \omega$.
Therefore, assuming all parameters are positive, the conditions for the existence of real and positive asymmetric fixed points are
\begin{equation}
    \omega \neq 0,
    \, \,
    \epsilon > \omega,
    \, \,
    \sigma > \gamma,
    \, \,
    \frac{(\sigma-\gamma)^{2}}{\omega^{2}}
    +
    \frac{4\delta}{\omega-\epsilon}
    >0.
\end{equation}

\subsubsection{Stability properties}

We now investigate the stability of the fixed points determined previously through a linear stability analysis. 
Considering Eqs.~(\ref{eq:6}), the linearized equations associated with this system can be written in the form $ \dot{\mathbf{q}} = \mathcal{J}(\phi,\psi)\mathbf{q}$, where $\mathcal{J}$ is the Jacobian matrix evaluated at the fixed points. 
The solution of this system can be written as $\mathbf{q}(\tau)=\mathbf{c}\,e^{\lambda\tau}$, where $\boldsymbol{c}$ is an eigenvector and $\lambda$ is the corresponding eigenvalue, determined from the characteristic polynomial
$   \lambda^{2} - \tr \mathcal{J}\,\lambda + \det \mathcal{J} = 0$,
where $\tr \mathcal{J}$ and $\det \mathcal{J}$ denote the trace and determinant of the Jacobian matrix, respectively.
The stability properties of the fixed points can therefore be determined from the eigenvalues \cite{silva2021basic}. 

Let us first analyze the stability of the asymmetric fixed points. Stability requires both eigenvalues to have negative real parts, which is guaranteed when
$\tr \mathcal{J} < 0$ and $\det \mathcal{J} > 0$.
In this case,
\begin{equation}
    \tr \mathcal{J}
    =
    - \frac{\epsilon (\sigma-\gamma)}{\omega},
    \qquad
    \det \mathcal{J}
    =
    -\omega (\omega - \epsilon )(\psi^*)^2.
\end{equation}
Using the existence and positivity conditions obtained previously, we conclude that the asymmetric fixed points are always stable.

Let us now analyze the symmetric fixed point, $(\psi^*,\phi_+)$. In this case, the stability analysis is simpler when performed directly from the eigenvalues. 
We obtain
\begin{equation}
    \lambda_{\psi}
    =
    (\sigma-\gamma) - \omega \phi^*,
\end{equation}
and
\begin{equation}
    \lambda_{\phi}
    =
    (\sigma-\gamma) - (\omega + \epsilon)\phi^*
    =
    -\sqrt{(\sigma-\gamma)^2 +4(\omega+\epsilon)\delta}.
\end{equation}
Therefore, we always have
$\lambda_{\phi}<0$.

For stability, it is necessary that $\lambda_{\psi}<0$. Expanding this condition yields
\begin{equation}
    \delta >
    \frac{(\epsilon - \omega)(\sigma-\gamma)^2}{4 \omega^{2}}.
\end{equation}
However, from the existence conditions we have
$\delta <
\frac{(\epsilon - \omega)(\sigma-\gamma)^2}{4 \omega^{2}}$.
Therefore, the symmetric fixed point is always a saddle point, which indicates that it is unstable.

\subsection{Adiabatic elimination}\label{subsec:AppB}

In this section, we derive the reduced Fokker--Planck equation used in the
main text. 
We focus on the regime in which the variable \(\phi\) relaxes much
faster than the chiral imbalance \(\psi\). In this case, \(\phi\) can be
treated as a fast variable, while \(\psi\) evolves on a slower time scale.

The full Fokker--Planck equation can be written in conservative form as
\begin{equation}\label{eq:B_1}
    \partial_{\tau}P(\phi,\psi,\tau)
    =
    - \partial_\phi J_{\phi}
    - \partial_\psi J_\psi,
\end{equation}
where \(J_{\phi}\) and \(J_{\psi}\) are the probability currents associated
with the variables \(\phi\) and \(\psi\), respectively.

The marginal probability density of the slow variable is defined as
\begin{equation}\label{eq:B_2}
    p(\psi,\tau)
    =
    \int d\phi\, P(\phi,\psi,\tau).
\end{equation}
Integrating Eq.~(\ref{eq:B_1}) over \(\phi\), we obtain
\begin{equation}\label{eq:B_3}
    \partial_{\tau} p(\psi,\tau)
    =
    - \partial_\psi \int d\phi\, J_\psi(\phi,\psi,\tau)
    - [J_\phi]_{\mathrm{boundary}}.
\end{equation}
For reflecting boundaries, or equivalently no probability flux in the
\(\phi\)-direction at the boundaries, the boundary term vanishes:
\[
[J_\phi]_{\mathrm{boundary}}=0.
\]
Therefore,
\begin{equation}\label{eq:B_4}
    \partial_\tau p(\psi,\tau)
    =
    - \partial_\psi j_\psi(\psi,\tau),
\end{equation}
where
\begin{equation}\label{eq:B_5}
    j_\psi(\psi,\tau)
    =
    \int d\phi\,J_{\psi}(\phi,\psi,\tau).
\end{equation}

The joint probability density can be written as
\begin{equation}\label{eq:B_6}
    P(\phi,\psi,\tau)
    =
    \rho(\phi|\psi,\tau)p(\psi,\tau),
\end{equation}
where \(\rho(\phi|\psi,\tau)\) is the conditional probability density of
\(\phi\) for a given value of \(\psi\). 
Under the adiabatic approximation,
the fast variable rapidly relaxes to a quasi-stationary conditional
distribution, so that
\begin{equation}\label{eq:B_7}
    P(\phi,\psi,\tau)
    \simeq
    \rho_{\mathrm{qs}}(\phi|\psi)p(\psi,\tau).
\end{equation}
At leading order in the time-scale separation, we neglect fluctuations of \(\phi\) around its
quasi-stationary value and approximate
\begin{equation}\label{eq:B_8}
    \rho_{\mathrm{qs}}(\phi|\psi)
    \simeq
    \delta(\phi-\phi^*(\psi)),
\end{equation}
where \(\phi^*(\psi)\) is the stable solution of
\(\alpha_\phi(\phi,\psi)=0\).

The effective drift and diffusion coefficients for the slow variable are then
obtained by averaging over the fast variable:
\begin{equation}\label{eq:B_9}
    a(\psi)
    =
    \int d\phi\,\alpha_\psi(\phi,\psi)\rho_{\mathrm{qs}}(\phi|\psi),
\end{equation}
and
\begin{equation}\label{eq:B_10}
    b(\psi)
    =
    \int d\phi\,\beta_{\psi\psi}(\phi,\psi)\rho_{\mathrm{qs}}(\phi|\psi).
\end{equation}
Using Eq. \eqref{eq:B_8}, these expressions reduce to
\begin{equation}
    a(\psi)
    =
    \alpha_\psi(\phi^*(\psi),\psi),
    \qquad
    b(\psi)
    =
    \beta_{\psi\psi}(\phi^*(\psi),\psi).
\end{equation}
Therefore, to leading order in the adiabatic approximation, the marginal
probability density \(p(\psi,\tau)\) satisfies the reduced Fokker--Planck
equation
\begin{equation}\label{eq:B_11}
    \partial_\tau p(\psi,\tau)
    =
    -\partial_\psi \left[a(\psi)p(\psi,\tau)\right]
    +
    \frac{1}{2}
    \partial^{2}_\psi\left[b(\psi)p(\psi,\tau)\right].
\end{equation}

The approximation above corresponds to retaining only the leading-order
contribution in the separation of time scales. In this limit, the conditional
distribution of the fast variable is approximated by a delta function centered
at the stable slow manifold \(\phi=\phi^*(\psi)\). Therefore, the finite width
of \(\rho_{\mathrm{qs}}(\phi|\psi)\), which is controlled by
\(\beta_{\phi\phi}\), is neglected. In addition, corrections associated with
the correlated fluctuations between the fast and slow variables, encoded in
\(\beta_{\phi\psi}\), are not retained. Thus, the reduced diffusion coefficient
is approximated by the direct slow component
\(\beta_{\psi\psi}(\phi^*(\psi),\psi)\).

\subsection{Splitting probability}\label{subsec:AppC}

In this section, we derive the backward equation for the splitting probability
$\Pi(\psi_0)$. 
Let the stochastic state variable $\psi$ evolve in the
interval $(\psi_-,\psi_+)$, where $\psi_-$ and $\psi_+$ are absorbing
boundaries.
We ask the following question: what is the probability that a system initially
at $\psi_0$ exits the interval through the upper boundary $\psi_+$ before reaching
the lower boundary $\psi_-$? We denote this probability by $\Pi(\psi_0)$.

Let $g(\psi_0,\tau)$ be the probability that a trajectory starting at $\psi_0$
exits through $\psi_+$ after time $\tau$. 
This probability is obtained by
integrating the outward probability flux through the upper boundary from
$\tau$ to $\infty$:
\begin{equation}\label{eq:C_1}
    g(\psi_0,\tau)
    =
    \int_{\tau}^{\infty}
    d\tau'\,
    j_{\psi_{0}}(\psi_+,\tau'|\psi_{0},0).
\end{equation}

The probability current is given by
\begin{equation}\label{eq:C_current}
    j_{\psi_{0}}(y,\tau'|\psi_0,0)
    =
    a(y)p(y,\tau'|\psi_0,0)
    -
    \frac{1}{2}
    \partial_y\left[
    b(y)p(y,\tau'|\psi_0,0)
    \right],
\end{equation}
evaluated at the absorbing boundary $y=\psi_+$. 
Here $y$ denotes the final
position, while $\psi_0$ denotes the initial position.

The backward operator acting on the initial condition $\psi_0$ is
\begin{equation}\label{eq:C_2}
    \mathcal{L}
    =
    a(\psi_0)\partial_{\psi_{0}}
    +
    \frac{1}{2}b(\psi_0)\partial_{\psi_{0}}^{2}.
\end{equation}
Applying this operator to Eq.~\eqref{eq:C_1}, we obtain
\begin{equation}\label{eq:C_3}
    \mathcal{L}g(\psi_0,\tau)
    =
    \int_{\tau}^{\infty}
    d\tau'\,
    \mathcal{L}
    j_{\psi_0}(\psi_+,\tau'|\psi_0,0).
\end{equation}

Since the current depends linearly on the transition probability density, and
the transition probability satisfies the backward Fokker--Planck equation, we
have
\begin{equation}\label{eq:C_4}
    \mathcal{L}
    j_{\psi_0}(\psi_+,\tau'|\psi_0,0)
    =
    \partial_{\tau'}
    j_{\psi_0}(\psi_+,\tau'|\psi_0,0).
\end{equation}
Therefore,
\begin{equation}\label{eq:C_5}
    \mathcal{L}g(\psi_0,\tau)
    =
    \int_{\tau}^{\infty}
    d\tau'\,
    \partial_{\tau'}
    j_{\psi_0}(\psi_+,\tau'|\psi_0,0).
\end{equation}
Assuming that the current vanishes at long times, this gives
\begin{equation}\label{eq:C_6}
    \mathcal{L}g(\psi_0,\tau)
    =
    -j_{\psi_0}(\psi_+,\tau|\psi_0,0).
\end{equation}
Differentiating Eq.~\eqref{eq:C_1} with respect to $\tau$
gives
\begin{equation}\label{eq:C_7}
    \partial_{\tau}g(\psi_0,\tau)
    =
    -j_{\psi_0}(\psi_+,\tau|\psi_0,0).
\end{equation}
Combining Eqs.~\eqref{eq:C_6} and \eqref{eq:C_7}, we obtain the backward
equation:
\begin{equation}\label{eq:C_8}
    \partial_{\tau}g(\psi,\tau)
    =
    \mathcal{L}g(\psi,\tau).
\end{equation}
The total probability of eventually exiting through the upper boundary is
obtained by taking $\tau=0$:
\begin{equation}\label{eq:C_9}
    \Pi(\psi_0)=g(\psi_0,0).
\end{equation}
If the initial condition lies inside the interval, $\psi_0\in(\psi_-,\psi_+)$,
there is no probability flux through the boundary at the initial time. Hence,
\begin{equation}
    j_{\psi_0}(\psi_+,0|\psi_0,0)=0,
\end{equation}
and therefore
\begin{equation}
    \partial_\tau g(\psi_0,0)=0.
\end{equation}
Evaluating Eq.~\eqref{eq:C_8} at $\tau=0$ gives
\begin{equation}\label{eq:C_11}
    a(\psi_0)\Pi'(\psi_0)
    +
    \frac{1}{2}b(\psi_0)\Pi''(\psi_0)
    =
    0.
\end{equation}

The equation above is solved with the absorbing boundary conditions
\begin{equation}
    \Pi(\psi_-)=0,
    \qquad
    \Pi(\psi_+)=1.
\end{equation}
These conditions follow directly from the definition of $\Pi(\psi_0)$: if the
system starts at $\psi_-$, the probability of exiting through $\psi_+$ is zero,
whereas if it starts at $\psi_+$, the exit through the upper boundary has
already occurred.

\subsection{Approximation of the splitting probability}\label{subsec:AppD}

In this section, we derive a first-order approximation for the splitting probability, given by Eq.~(\ref{eq:24}), in the vicinity of $\psi_0=0$. Expanding $\Pi(\psi_0)$ about $\psi_0=0$, we obtain
\begin{equation}
\Pi(\psi_0)
\simeq
\Pi(0)
+
\Pi'(0)\psi_0
+
\mathcal{O}(\psi_0^2).
\end{equation}
Assuming the symmetry of the problem implies $\Pi(0)=1/2$, and using Eq.~(\ref{eq:24}) together with the fundamental theorem of calculus, we find
\begin{equation}
\Pi'(0)
=
\left(
\int_{\psi_-}^{\psi_+}
\exp[\tilde{\Phi}(s)]ds
\right)^{-1}.
\end{equation}
Therefore,
\begin{equation}
\Pi(\psi_0)
\simeq
\frac{1}{2}
+
\left(
\int_{\psi_-}^{\psi_+}
\exp[\tilde{\Phi}(s)]ds
\right)^{-1}
\psi_0
+
\mathcal{O}(\psi_0^2),
\end{equation}
where
\begin{equation}
\tilde{\Phi}(s)
=
-
\int_0^s
\frac{2a(u)}{b(u)}
du.
\end{equation}

To approximate $\tilde{\Phi}(s)$ near $s=0$, we use the fact that $a(0)=0$ and that $a$ and $b$ are sufficiently smooth, with $b(0)\neq 0$. Expanding the integrand to leading order gives
\begin{equation}
\frac{2a(u)}{b(u)}
\simeq
\frac{2a'(0)}{b(0)}u.
\end{equation}
Hence,
\begin{equation}
\tilde{\Phi}(s)
\simeq
-
\frac{a'(0)}{b(0)}s^2
=
-\kappa s^2,
\qquad
\kappa
=
\frac{a'(0)}{b(0)}.
\end{equation}

Provided that this local quadratic approximation is valid over the relevant integration region, or that the integral is dominated by values of $s$ close to zero, we obtain
\begin{equation}
\Pi(\psi_0)
\simeq
\frac{1}{2}
+
\left(
\int_{\psi_-}^{\psi_+}
\exp[-\kappa s^2]ds
\right)^{-1}
\psi_0
+
\mathcal{O}(\psi_0^2).
\end{equation}

Finally, using the symmetry relation $\psi_-=-\psi_+$ and assuming $\kappa>0$, the Gaussian integral becomes
\begin{equation}
\int_{-\psi_+}^{\psi_+}
\exp[-\kappa s^2]ds
=
\sqrt{\frac{\pi}{\kappa}}
\erf
\left(
\psi_+\sqrt{\kappa}
\right).
\end{equation}
Thus, the first-order approximation for the splitting probability is
\begin{equation}
\Pi(\psi_0)
\simeq
\frac{1}{2}
+
\frac{\sqrt{\kappa}}
{\sqrt{\pi}
\erf
\left(
\psi_+\sqrt{\kappa}
\right)}
\psi_0
+
\mathcal{O}(\psi_0^2).
\end{equation}

\subsection{Mean first-passage time}\label{subsec:AppE}

In this section we derive the equation satisfied by the mean first-passage time, $T(\psi_0)$~\cite{gardiner2009stochastic}.
Assume that the stochastic state variable is initially located in the interval $(\psi_-,\psi_+)$ and evolves in time until it reaches one of the absorbing boundaries $\psi_-$ or $\psi_+$. 
We begin by defining the survival probability, i.e., the probability that at time $t$ the system remains inside the interval $(\psi_-,\psi_+)$,
\begin{equation}
G(\psi_0,t)
=
\int_{\psi_-}^{\psi_+}
p(\psi,t|\psi_0,0)\,d\psi,
\end{equation}
where $p(\psi,t|\psi_0,0)$ denotes the transition probability density.

Since the process is time-homogeneous,
\[
p(\psi,t|\psi_0,0)
=
p(\psi,0|\psi_0,-t),
\]
and the transition density satisfies the backward Fokker--Planck equation,
\begin{equation}
\partial_\tau p
=
\mathcal{L}p,
\end{equation}
where the operator $\mathcal{L}$ is given by Eq.~(\ref{eq:C_2}). Integrating this equation over the interval $(\psi_-,\psi_+)$ and noting that the operator acts on the initial condition $\psi_0$, one finds that the survival probability also satisfies the backward equation,
\begin{equation}
\partial_\tau G(\psi_0,\tau)
=
\mathcal{L}G(\psi_0,\tau).
\end{equation}

The survival probability satisfies the initial condition $G(\psi_0,0)=1$,
for $\psi_0\in(\psi_-,\psi_+)$, and the boundary conditions $G(\psi_-,t)=G(\psi_+,t)=0.$
Since $G(\psi_0,t)=\mathrm{Prob}(T\ge t)$, the mean of any function of the first-passage time is
\begin{equation}
\langle f(T)\rangle
=
-\int_0^\infty
f(\tau)\,
\partial_\tau G(\psi_0,\tau)
\,d\tau.
\end{equation}
In particular, choosing $f(T)=T$ and integrating by parts yields
\begin{equation}
\langle T\rangle
=
\int_0^\infty
G(\psi_0,\tau)\,d\tau .
\end{equation}
Finally, integrating the backward equation over the interval $\tau\in(0,\infty)$ gives
\begin{equation}
a(\psi_0)\,T'(\psi_0)
+
\frac12 b(\psi_0)\,T''(\psi_0)
=
-1,
\end{equation}
subject to the absorbing boundary conditions
\begin{equation}
T(\psi_-)=0,
\qquad
T(\psi_+)=0.
\end{equation}

\subsection{Saddle-point approximation}\label{subsec:AppF}

Here, we aim to obtain the approximate escape rate (Eq.~\eqref{eq:33}). 
It is useful first to rewrite the full solution in terms of the effective potential, using Eq.~\eqref{eq:19}, namely,
\begin{equation}
\Phi(\psi) = U(\psi) - \log b(\psi).
\end{equation}
Substituting this expression into Eq.~(\ref{eq:32}), we obtain
\begin{equation}
T(\psi_-) = 2 \int_{\psi_-}^{0} ds  \frac{\exp[U(s)]}{b(s)}
\int_{\psi_-}^{s} du  \exp[-U(u)].
\end{equation}

As can be observed from Fig.~\ref{fig:2}, the following properties hold for our problem:
\begin{equation}
U'(\psi_-)=0,
\quad
U'(0)=0,
\quad
U''(\psi_-)>0,
\quad
U''(0)<0.
\end{equation}
In other words, $U(\psi)$ has a minimum at $\psi=\psi_-$ and a maximum at $\psi=0$. 
Therefore, the outer integral is dominated by the vicinity of the maximum point, whereas the inner integral is dominated by the vicinity of the minimum. 
Under these conditions, and assuming the potential barrier is sufficiently high, we can apply the saddle-point approximation.

Close to $\psi=\psi_-$, we have
$U(\psi) \simeq U(\psi_-)+
 \frac{1}{2}U''(\psi_-)(\psi-\psi_-)^{2}$.
  Similarly, close to $\psi=0$, we have
  $U(\psi) \simeq U(0)
+ \frac{1}{2}U''(0)\psi^{2}$.
Using these approximations in the two integrals gives
\begin{equation}
\begin{split}
\int_{\psi_-}^{s} du \exp[-U(u)]
\simeq
\exp[-U(\psi_-)]
\sqrt{\frac{\pi}{2U''(\psi_-)}} ,
\end{split}
\end{equation}
and
\begin{equation}
\int_{\psi_-}^{0} ds  \frac{\exp[U(s)]}{b(s)}
\simeq
\frac{\exp[U(0)]}{b(0)}
\sqrt{\frac{\pi}{2|U''(0)|}} .
\end{equation}
Combining both results, we obtain
\begin{equation}
T(\psi_-)
\simeq
\frac{\pi}{b(0)\sqrt{U''(\psi_-)|U''(0)|}}
\exp[\Delta U],
\end{equation}
where
$\Delta U = U(0)-U(\psi_-)$.
Therefore, since $k=1/T(\psi_-)$, the escape rate is
\begin{equation}
k
\simeq
\frac{b(0)\sqrt{U''(\psi_-)|U''(0)|}}{\pi}
\exp[-\Delta U],
\end{equation}
as we wanted to show.

\bibliography{ref.bib}

@article{gillespie1977exact,
  title={Exact stochastic simulation of coupled chemical reactions},
  author={Gillespie, Daniel T},
  journal={The journal of physical chemistry},
  volume={81},
  number={25},
  pages={2340--2361},
  year={1977},
  publisher={ACS Publications}
}

@book{gardiner2009stochastic,
  title={Stochastic methods},
  author={Gardiner, Crispin},
  volume={4},
  year={2009},
  publisher={Springer Berlin Heidelberg}
}

@article{sugimori2008fluctuation,
  title={Fluctuation induced homochirality},
  author={Sugimori, Takeshi and Hyuga, Hiroyuki and Saito, Yukio},
  journal={Journal of the Physical Society of Japan},
  volume={77},
  number={6},
  pages={064606--064606},
  year={2008},
  publisher={The Physical Society of Japan}
}

@article{boscheto2015spontaneous,
  title={Spontaneous chiral symmetry breaking for finite systems},
  author={Boscheto, Emerson and L{\'o}pez-Castillo, Alejandro},
  journal={ChemPhysChem},
  volume={16},
  number={17},
  pages={3728--3735},
  year={2015},
  publisher={Wiley Online Library}
}

@article{plasson2004recycling,
  title={Recycling Frank: spontaneous emergence of homochirality in noncatalytic systems},
  author={Plasson, Rapha{\"e}l and Bersini, Hugues and Commeyras, Auguste},
  journal={Proceedings of the National Academy of Sciences},
  volume={101},
  number={48},
  pages={16733--16738},
  year={2004},
  publisher={National Academy of Sciences}
}

@article{blackmond2008re,
  title={Re-examination of reversibility in reaction models for the spontaneous emergence of homochirality},
  author={Blackmond, Donna G and Matar, Omar K},
  journal={The Journal of Physical Chemistry B},
  volume={112},
  number={16},
  pages={5098--5104},
  year={2008},
  publisher={ACS Publications}
}

@article{saito2004complete,
  title={Complete homochirality induced by nonlinear autocatalysis and recycling},
  author={Saito, Yukio and Hyuga, Hiroyuki},
  journal={Journal of the Physical Society of Japan},
  volume={73},
  number={1},
  pages={33--35},
  year={2004},
  publisher={The Physical Society of Japan}
}

@article{kondepudi1993kinetics,
  title={Kinetics of chiral symmetry breaking in crystallization},
  author={Kondepudi, Dilip K and Bullock, Kim L and Digits, Jennifer A and Hall, John K and Miller, Jason M},
  journal={Journal of the American Chemical Society},
  volume={115},
  number={22},
  pages={10211--10216},
  year={1993},
  publisher={ACS Publications}
}

@article{gillet2024differences,
  title={Differences in enantiomeric diffusion can lead to selective chiral amplification},
  author={Gillet, Jean and Geerts, Yves and Rongy, Laurence and De Decker, Yannick},
  journal={Proceedings of the National Academy of Sciences},
  volume={121},
  number={17},
  pages={e2319770121},
  year={2024},
  publisher={National Academy of Sciences}
}

@book{atkins1999chemical,
  title={Chemical principles: The quest for insight},
  author={Atkins, Peter William and Jones, Loretta and Laverman, Leroy and Patterson, James and Young, Kelley},
  volume={4},
  year={1999},
  publisher={WH Freeman}
}

@article{silva2021basic,
  title={Basic concepts of self-organized phenomena in chemical systems},
  author={Silva-Dias, Leonardo},
  journal={Qu{\'\i}mica Nova},
  volume={44},
  pages={646--654},
  year={2021},
  publisher={SciELO Brasil}
}

@article{hochberg2023chiral,
  title={Chiral selectivity vs. noise in spontaneous mirror symmetry breaking},
  author={Hochberg, David and Buhse, Thomas and Micheau, Jean-Claude and Rib{\'o}, Josep M},
  journal={Physical Chemistry Chemical Physics},
  volume={25},
  number={46},
  pages={31583--31595},
  year={2023},
  publisher={The Royal Society of Chemistry}
}

@article{soai2014asymmetric,
  title={Asymmetric autocatalysis of pyrimidyl alkanol and its application to the study on the origin of homochirality},
  author={Soai, Kenso and Kawasaki, Tsuneomi and Matsumoto, Arimasa},
  journal={Accounts of chemical research},
  volume={47},
  number={12},
  pages={3643--3654},
  year={2014},
  publisher={ACS Publications}
}

@article{sato2003amplification,
  title={Amplification of chirality from extremely low to greater than 99.5\% ee by asymmetric autocatalysis},
  author={Sato, Itaru and Urabe, Hiroki and Ishiguro, Saori and Shibata, Takanori and Soai, Kenso},
  journal={Angewandte Chemie-International Edition},
  volume={42},
  number={3},
  pages={315--317},
  year={2003},
  publisher={John Wiley and Sons Ltd}
}

@article{soai1995asymmetric,
  title={Asymmetric autocatalysis and amplification of enantiomeric excess of a chiral molecule},
  author={Soai, Kenso and Shibata, Takanori and Morioka, Hiroshi and Choji, Kaori},
  journal={Nature},
  volume={378},
  number={6559},
  pages={767--768},
  year={1995}
}

@article{xu2025ubiquitous,
  title={Ubiquitous chiral symmetry breaking of conjugated polymers via liquid--liquid phase separation},
  author={Xu, Zhuang and Wang, Ziming and Zhang, Rui and Verma, Archana and Neu, Justin S and Bairagya, Priyotosh and Jeon, Sanghyun and Damron, Alec M and Rinehart, Joshua M and Wagner, Justine S and others},
  journal={Journal of the American Chemical Society},
  volume={147},
  number={37},
  pages={33530--33544},
  year={2025},
  publisher={ACS Publications}
}

@article{sato2001evidence,
  title={Evidence of asymmetric autocatalysis in the enantioselective addition of diisopropylzinc to pyrimidine-5-carbaldehyde using chiral pyrimidyl alkanol},
  author={Sato, Itaru and Omiya, Daisuke and Tsukiyama, Koichi and Ogi, Yoshihiro and Soai, Kenso},
  journal={Tetrahedron: Asymmetry},
  volume={12},
  number={14},
  pages={1965--1969},
  year={2001},
  publisher={Elsevier}
}

@article{tsogoeva2009generation,
  title={Generation of highly enantioenriched crystalline products in reversible asymmetric reactions with racemic or achiral catalysts},
  author={Tsogoeva, Svetlana B and Wei, Shengwei and Freund, Matthias and Mauksch, Michael},
  journal={Angewandte Chemie},
  volume={121},
  number={3},
  pages={598--602},
  year={2009},
  publisher={Wiley Online Library}
}

@article{viedma2008evolution,
  title={Evolution of solid phase homochirality for a proteinogenic amino acid},
  author={Viedma, Cristobal and Ortiz, Jos{\'e} E and Torres, Trinidad de and Izumi, Toshiko and Blackmond, Donna G},
  journal={Journal of the American Chemical Society},
  volume={130},
  number={46},
  pages={15274--15275},
  year={2008},
  publisher={ACS Publications}
}

@article{noorduin2008emergence,
  title={Emergence of a single solid chiral state from a nearly racemic amino acid derivative},
  author={Noorduin, Wim L and Izumi, Toshiko and Millemaggi, Alessia and Leeman, Michel and Meekes, Hugo and Van Enckevort, Willem JP and Kellogg, Richard M and Kaptein, Bernard and Vlieg, Elias and Blackmond, Donna G},
  journal={Journal of the American Chemical Society},
  volume={130},
  number={4},
  pages={1158--1159},
  year={2008},
  publisher={ACS Publications}
}

@article{cartwright2004chiral,
  title={Chiral symmetry breaking during crystallization: An advection-mediated nonlinear autocatalytic process},
  author={Cartwright, Julyan HE and Garc{\'\i}a-Ruiz, Juan Manuel and Piro, Oreste and Sainz-D{\'\i}az, C Ignacio and Tuval, Idan},
  journal={Physical review letters},
  volume={93},
  number={3},
  pages={035502},
  year={2004},
  publisher={APS}
}

@article{buhse2000chiral,
  title={Chiral symmetry breaking in crystallization: The role of convection},
  author={Buhse, T and Durand, D and Kondepudi, D and Laudadio, J and Spilker, S},
  journal={Physical Review Letters},
  volume={84},
  number={19},
  pages={4405},
  year={2000},
  publisher={APS}
}

@book{nicolis1977self,
  title={Self-Organization in Nonequilibrium Systems: From Dissipative Structures to Order Through Fluctuations},
  author={Nicolis, Grégoire and Prigogine, Ilya},
  year={1977},
  publisher={Wiley},
  address={New York},
  isbn={9780471024019}
}

@article{hegstrom1990handedness,
  title={The handedness of the universe},
  author={Hegstrom, Roger A and Kondepudi, Dilip K},
  journal={Scientific American},
  volume={262},
  number={1},
  pages={108--115},
  year={1990},
  publisher={JSTOR}
}

@article{viedma2005chiral,
  title={Chiral Symmetry Breaking During Crystallization: Complete Chiral Purity<? format?> Induced by Nonlinear Autocatalysis and Recycling},
  author={Viedma, Cristobal},
  journal={Physical review letters},
  volume={94},
  number={6},
  pages={065504},
  year={2005},
  publisher={APS}
}

@article{kondepudi1990chiral,
  title={Chiral symmetry breaking in sodium chlorate crystallization},
  author={Kondepudi, Dilip K and Kaufman, Rebecca J and Singh, Nolini},
  journal={Science},
  volume={250},
  number={4983},
  pages={975--976},
  year={1990},
  publisher={American Association for the Advancement of Science}
}

@article{kondepudi2001chiral,
  title={Chiral autocatalysis, spontaneous symmetry breaking, and stochastic behavior},
  author={Kondepudi, Dilip K and Asakura, Kouichi},
  journal={Accounts of chemical research},
  volume={34},
  number={12},
  pages={946--954},
  year={2001},
  publisher={ACS Publications}
}

@article{soai2019asymmetric,
  title={Asymmetric autocatalysis. Chiral symmetry breaking and the origins of homochirality of organic molecules},
  author={Soai, Kenso},
  journal={Proceedings of the Japan Academy, Series B},
  volume={95},
  number={3},
  pages={89--110},
  year={2019},
  publisher={The Japan Academy}
}

@article{silva2018spontaneous,
  title={Spontaneous symmetry breaking of population: Stochastic Lotka--Volterra model for competition among two similar preys and predators},
  author={Silva-Dias, L and L{\'o}pez-Castillo, A},
  journal={Mathematical biosciences},
  volume={300},
  pages={36--46},
  year={2018},
  publisher={Elsevier}
}

@article{kondepudi1984chiral,
  title={Chiral-symmetry-breaking states and their sensitivity in nonequilibrium chemical systems},
  author={Kondepudi, Dilip K and Nelson, George W},
  journal={Physica A: Statistical Mechanics and its Applications},
  volume={125},
  number={2-3},
  pages={465--496},
  year={1984},
  publisher={Elsevier}
}

@article{kondepudi1983chiral,
  title={Chiral symmetry breaking in nonequilibrium systems},
  author={Kondepudi, D\_K\_ and Nelson, G-W\_},
  journal={Physical Review Letters},
  volume={50},
  number={14},
  pages={1023},
  year={1983},
  publisher={APS}
}

@article{frey2010evolutionary,
  title={Evolutionary game theory: Theoretical concepts and applications to microbial communities},
  author={Frey, Erwin},
  journal={Physica A: Statistical Mechanics and its Applications},
  volume={389},
  number={20},
  pages={4265--4298},
  year={2010},
  publisher={Elsevier}
}

@article{reichenbach2006coexistence,
  title={Coexistence versus extinction in the stochastic cyclic Lotka-Volterra model},
  author={Reichenbach, Tobias and Mobilia, Mauro and Frey, Erwin},
  journal={Physical Review E—Statistical, Nonlinear, and Soft Matter Physics},
  volume={74},
  number={5},
  pages={051907},
  year={2006},
  publisher={APS}
}

@article{kondepudi1985weak,
  title={Weak neutral currents and the origin of biomolecular chirality},
  author={Kondepudi, Dilip K and Nelson, George W},
  journal={Nature},
  volume={314},
  number={6010},
  pages={438--441},
  year={1985},
  publisher={Nature Publishing Group UK London}
}

@article{lente2009stochastic,
  title={Stochastic kinetic analysis of the Frank model. Stochastic approach to flow-through reactors},
  author={Lente, G{\'a}bor and Ditr{\'o}i, Tam{\'a}s},
  journal={The Journal of Physical Chemistry B},
  volume={113},
  number={20},
  pages={7237--7242},
  year={2009},
  publisher={ACS Publications}
}

@article{lente2005stochastic,
  title={Stochastic kinetic models of chiral autocatalysis: a general tool for the quantitative interpretation of total asymmetric synthesis},
  author={Lente, G{\'a}bor},
  journal={The Journal of Physical Chemistry A},
  volume={109},
  number={48},
  pages={11058--11063},
  year={2005},
  publisher={ACS Publications}
}

@article{saito2013colloquium,
  title={Colloquium: Homochirality: Symmetry breaking in<? format?> systems driven far from equilibrium},
  author={Saito, Yukio and Hyuga, Hiroyuki},
  journal={Reviews of Modern Physics},
  volume={85},
  number={2},
  pages={603--621},
  year={2013},
  publisher={APS}
}

@article{de2025stochastic,
  title={Stochastic closed Frank model in two dimensions: Chiral symmetry breaking driven by diffusive control over bounded surfaces},
  author={de Souza Filho, Jos{\'e} C{\^a}ndido and L{\'o}pez-Castillo, Alejandro},
  journal={Physical Review E},
  volume={112},
  number={5},
  pages={054114},
  year={2025},
  publisher={APS}
}

@article{pineros2022spontaneous,
  title={Spontaneous chiral symmetry breaking in a random driven chemical system},
  author={Pi{\~n}eros, William D and Tlusty, Tsvi},
  journal={Nature Communications},
  volume={13},
  number={1},
  pages={2244},
  year={2022},
  publisher={Nature Publishing Group UK London}
}

@article{deng2024symmetry,
  title={Symmetry breaking and chiral amplification in prebiotic ligation reactions},
  author={Deng, Min and Yu, Jinhan and Blackmond, Donna G},
  journal={Nature},
  volume={626},
  number={8001},
  pages={1019--1024},
  year={2024},
  publisher={Nature Publishing Group UK London}
}

@article{jafarpour2017noise,
  title={Noise-induced symmetry breaking far from equilibrium and the emergence of biological homochirality},
  author={Jafarpour, Farshid and Biancalani, Tommaso and Goldenfeld, Nigel},
  journal={Physical Review E},
  volume={95},
  number={3},
  pages={032407},
  year={2017},
  publisher={APS}
}

@article{frank1953spontaneous,
  title={On spontaneous asymmetric synthesis},
  author={Frank, Frederic C},
  journal={Biochimica et biophysica acta},
  volume={11},
  pages={459--463},
  year={1953},
  publisher={Elsevier}
}

@article{jafarpour2015noise,
  title={Noise-induced mechanism for biological homochirality of early life self-replicators},
  author={Jafarpour, Farshid and Biancalani, Tommaso and Goldenfeld, Nigel},
  journal={Physical review letters},
  volume={115},
  number={15},
  pages={158101},
  year={2015},
  publisher={APS}
}

@article{lente2010role,
  title={The role of stochastic models in interpreting the origins of biological chirality},
  author={Lente, G{\'a}bor},
  journal={Symmetry},
  volume={2},
  number={2},
  pages={767--798},
  year={2010},
  publisher={Molecular Diversity Preservation International}
}

@article{silva2017stochastic,
  title={Stochastic chiral symmetry breaking process besides the deterministic one},
  author={Silva-Dias, L and Lopez-Castillo, A},
  journal={Physical Chemistry Chemical Physics},
  volume={19},
  number={43},
  pages={29424--29428},
  year={2017},
  publisher={Royal Society of Chemistry}
}

@article{piaggi2023critical,
  title={Critical behavior in a chiral molecular model},
  author={Piaggi, Pablo M and Car, Roberto and Stillinger, Frank H and Debenedetti, Pablo G},
  journal={The Journal of Chemical Physics},
  volume={159},
  number={11},
  year={2023},
  publisher={AIP Publishing}
}

\newpage

\end{document}